\documentclass[12pt]{article}

\usepackage[a4paper,pdftex]{geometry}
\usepackage[T1]{fontenc}
\usepackage[english]{babel}
\usepackage[authoryear]{natbib}
\usepackage[small]{titlesec}
\usepackage[justification=centering]{caption}
\usepackage{hyperref}

\usepackage{amsmath,amsthm,amssymb,graphicx,enumerate,booktabs,bigstrut,rotating,multirow,float,etoolbox,lmodern,comment,listings,qtree,a4wide,moresize,bbm,subcaption,tikz,pdfescape,mathtools,flafter,enumitem,setspace,ragged2e,colortbl,lineno,lscape,pdflscape,stackengine,pgffor}

\usetikzlibrary{decorations.pathreplacing,positioning, arrows.meta,shapes}

\newcolumntype{L}[1]{>{\raggedright\let\newline\\\arraybackslash\hspace{0pt}}m{#1}}
\newcolumntype{C}[1]{>{\centering\let\newline\\\arraybackslash\hspace{0pt}}m{#1}}
\newcolumntype{R}[1]{>{\raggedleft\let\newline\\\arraybackslash\hspace{0pt}}m{#1}}

\hypersetup{colorlinks,linkcolor={red},citecolor={blue},urlcolor={blue}}

\makeatletter
\renewcommand\subsubsection{\@startsection{subsubsection}{3}{\z@}%
	{-3.25ex\@plus -1ex \@minus -.2ex}%
	{-1.5ex \@plus -.2ex}%
	{\normalfont\normalsize\bfseries}
}
\def\@biblabel#1{\hspace*{-\labelsep}}

\makeatother

\usepackage{datatool}
\DTLsetseparator{,}
\DTLloaddb[keys={key}]{est}{results/kv_store.csv}
\newcommand{\getval}[1]{\DTLfetch{est}{key}{#1}{value}}

\def\sym#1{\ifmmode^{#1}\else\(^{#1}\)\fi}

\definecolor{green}{RGB}{0,158,115}
\definecolor{yellow}{RGB}{240,228,66}

\pdfpageheight\paperheight
\pdfpagewidth\paperwidth

\begin{document}
\begin{filecontents*}{results/kv_store.csv}
key,value,timestamp,label
forgone_mean_mistake,"18,453",,Mean foregone bonus | mistake ($/yr)
forgone_mean_p90,"58,138",,"Mean foregone bonus, top decile of mistakes ($/yr)"
bonus_mean_annual,"298,265",,Mean annual bonus per dealer-year ($)
xsec_effect_lo_pct,8,,"X-sec effect, low end (% lower)"
xsec_effect_hi_pct,20,,"X-sec effect, high end (% lower)"
dma_gap_pct,11,,Most- vs least-competitive DMA mistake-rate gap (pp)
within_effect_lo,2,,"Within effect, low end (pp/SD)"
within_effect_hi,6,,"Within effect, high end (pp/SD)"
within_top3_pval,0.10,,"Within-dealer p-value, top-3 share (Panel B)"
within_hhi_pval,0.13,,"Within-dealer p-value, HHI (Panel B)"
pred_rate_perfcomp_pct,16,,"Predicted mistake rate, perfect competition (%)"
pred_rate_monopoly_pct,34,,"Predicted mistake rate, perfect monopoly (%)"
mistake_rate_q4_pct,17,,"Mistake rate, most-competitive quartile (%)"
forgone_mean_q4,"18,231",,"Mean foregone value, most-competitive quartile ($)"
within_top1_coef,0.176,,Within coef top1 (level)
yrfe_pp_lo,10.9,,"Year-FE perfect-comp vs monopoly gap, min over measures (pp)"
yrfe_pp_hi,17.8,,"Year-FE perfect-comp vs monopoly gap, max over measures (pp)"
yrfe_rel_lo,61,,"Year-FE effect relative to constant, min over measures (pct)"
yrfe_rel_hi,163,,"Year-FE effect relative to constant, max over measures (pct)"
within_pp_lo,14.7,,"Within perfect-comp vs monopoly gap, min over measures (pp)"
within_pp_hi,25.3,,"Within perfect-comp vs monopoly gap, max over measures (pp)"
comp_hhi_mean_dma,0.120,,"Mean HHI, Panel D"
comp_top1_mean_pct,21.4,,"Mean top-1 share, Panel D (pct)"
comp_top3_mean_pct,40.7,,"Mean top-3 share, Panel D (pct)"
comp_top5_mean_pct,52.6,,"Mean top-5 share, Panel D (pct)"
binsc_xlo_pct,4,,Binscatter low-end top1 share (pct)
binsc_xhi_pct,72,,Binscatter high-end top1 share (pct)
binsc_lo_pct,11.6,,"Fitted mistake rate, binscatter low end (pct)"
binsc_hi_pct,22.9,,"Fitted mistake rate, binscatter high end (pct)"
binsc_gap_pp,11.3,,Binscatter low->high mistake-rate gap (pp)
mistake_rate_pct,20,,Overall mistake rate = Table 2 col 1 (%)
mistake_low_pct,15.9,,"Suboptimal rate, low-tier enrollees (pct)"
mistake_high_pct,4.2,,"Suboptimal rate, high-tier enrollees (pct)"
mistake_tier_gap_pp,12,,Low- minus high-tier gap (pp)
mistake_should_high_pct,11.9,,Should have chosen high tier (pct)
mistake_should_low_pct,0.9,,Should have chosen low tier (pct)
mistake_program_pct,7.3,,Should have forgone program (pct)
mistake_program_high_pct,3.4,,"Should have forgone program, high-tier choosers, Tab2 col2 (pct)"
mistake_program_low_pct,4.0,,"Should have forgone program, low-tier choosers, Tab2 col3 (pct)"
shr_noenroll_better_pct,7.7,,"Share better off not enrolling, Fig 2A (pct)"
shr_wrongtier_pct,16.1,,"Share chose suboptimal tier, Fig 2B (pct)"
noexit_top1_coef,0.184,,"No-exit within coef top1, dealer+yr FE (level)"
n_exit_dealers,461,,N dealers that exit (Table 6 sample)
exit_share_pct,11.1,,Exiting dealers as pct of sample
exit_xsec_pp,0.6,,"Largest cross-sectional exit-on-concentration effect (pp), tab:exit Panel A"
exit_within_pp,3.5,,"Largest within-dealer exit-on-concentration effect (pp), tab:exit Panel B"
n_dealers,"4,170",,N incumbent dealerships (analysis sample)
n_dealer_months,"150,096",,N dealership-month observations (analysis sample)
n_changed_tier,24,,N dealership-months dropped for mid-year tier switch
mistake_exp_pct,20,,"Mistake rate, experienced (pct)"
mistake_new_pct,33,,"Mistake rate, new entrants (pct)"
mistake_exp_gap_pp,13,,New minus experienced gap (pp)
mistake_exp_relinc_pct,65,,"Relative increase, new vs experienced (pct)"
exp_lowtier_share_pct,79,,Experienced: pct of mistakes that are low-tier
new_hightier_share_pct,53,,New: pct of mistakes that are over-investment
sumstat_employees,7.9,,"Mean salespeople, Panel A"
sumstat_sales,60.0,,"Mean vehicle sales/yr, Panel A"
sumstat_truck_pct,84.1,,"Truck share, Panel A (pct)"
sumstat_enroll_pct,59.5,,"Pct enrolled >=1 year, Panel A"
sumstat_years_enrolled,1.6,,"Mean years enrolled, Panel A"
sumstat_sales_enrolled,77.9,,"Mean sales/yr, enrolled, Panel B"
sumstat_tier1_pct,81.4,,"Pct enrolled choosing Tier 1, Panel B"
sumstat_same_enroll_pct,94.3,,"Pct maintaining enrollment status, Panel C"
sumstat_stay_enrolled_pct,52.2,,"Pct remaining enrolled, Panel C"
sumstat_stay_unenrolled_pct,42.0,,"Pct remaining non-enrolled, Panel C"
sumstat_change_enroll_pct,5.7,,"Pct changing enrollment status, Panel C"
sumstat_enter_enroll_pct,3.9,,"Pct non-enrolled to enrolled, Panel C"
sumstat_exit_enroll_pct,1.8,,"Pct enrolled to non-enrolled, Panel C"
sumstat_same_tier_pct,88.6,,"Pct maintaining tier selection, Panel C"
sched_t1_b2,100,,"Tier1 bonus per vehicle, 95-100 pct of goal (2017)"
sched_t1_b3,200,,"Tier1 bonus per vehicle, 100-106 pct (2017)"
sched_t1_b4,500,,"Tier1 bonus per vehicle, 106-120 pct (2017)"
sched_t1_b5,600,,"Tier1 bonus per vehicle, over 120 pct (2017)"
sched_t2_b2,50,,"Tier2 bonus per vehicle, 95-100 pct (2017)"
sched_t2_b3,150,,"Tier2 bonus per vehicle, 100-106 pct (2017)"
sched_t2_b4,350,,"Tier2 bonus per vehicle, 106-120 pct (2017)"
sched_t2_b5,400,,"Tier2 bonus per vehicle, over 120 pct (2017)"
cost_wrongtier,"19,003",,Mean wrong-tier cost among wrong-tier participants ($/yr)
median_annual_sales,538,,"Median annual sales, enrolled dealer-years"
growth_gt10_pct,18.5,,Pct dealer-years YoY sales growth over 10pct
growth_gt30_pct,4.3,,Pct dealer-years YoY sales growth over 30pct
growth_gt50_pct,1.5,,Pct dealer-years YoY sales growth over 50pct
shrink_gt10_pct,31.0,,Pct dealer-years YoY sales shrinkage over 10pct
shrink_gt14_pct,21.5,,Pct dealer-years YoY sales shrinkage over 14pct
shrink_gt27_pct,4.8,,Pct dealer-years YoY sales shrinkage over 27pct
yoy_n_dealeryears,"7,619",,"YoY sample dealer-years, fig:yoy_sales"
yoy_n_dealers,"3,910",,"YoY sample dealers, fig:yoy_sales"
yoy_units_lo,-552,,"YoY units change p1 (winsor low), fig:yoy_sales"
yoy_units_hi,450,,"YoY units change p99 (winsor high), fig:yoy_sales"
yoy_pct_lo,-42.6,,"YoY percent change p1 (winsor low), fig:yoy_sales"
yoy_pct_hi,58.8,,"YoY percent change p99 (winsor high), fig:yoy_sales"
re_overinvest_growth_pct,30,,"Req sales growth to justify high tier, over-invest bin (pct, no-filter engine median)"
re_underinvest_shrink_pct,14,,"Req sales shrinkage to justify low tier, under-invest bin (pct, no-filter engine median)"
re_underinvest_shrink_far_pct,27,,"Req sales shrinkage to justify low tier, far-up 50-60k bin (pct, no-filter engine median)"
choice_high_lowinc_pct,50,,"High-tier choice rate in lowest diff-profit bin, fig:choice_tier1_profit (pct)"
\end{filecontents*}

	
	\title{Competition and Anomalies Redux:\\ Evidence from U.S. Auto Dealers\thanks{\noindent Huffman: Cornell University, SC Johnson College of Business and IZA (\href{mailto:dbh234@cornell.edu }{dbh234@cornell.edu}), Pierce: Washington University in St. Louis (\href{mailto:pierce@wustl.edu}{pierce@wustl.edu}), Rees-Jones: University of Pennsylvania, Wharton School and NBER (\href{mailto:alre@wharton.upenn.edu}{alre@wharton.upenn.edu}),  
			Reyes: Middlebury College and IZA (\href{mailto:greyes@middlebury.edu}{greyes@middlebury.edu}). 
			We are grateful to Maritz, LLC for providing the data necessary for this project. No authors accepted compensation from any party for conducting this research. We thank Zamir Ticknor for excellent research assistance. }}
	\renewcommand{\today}{\ifcase \month \or January\or February\or March\or %
		April\or May \or June\or July\or August\or September\or October\or November\or %
		December\fi \ \number \year} 
	\date{\today}
	
	\author{David Huffman \and Lamar Pierce \and Alex Rees-Jones \and Germán Reyes}

	\maketitle
	
	\begin{abstract} 	
		\begin{singlespace}
			\noindent 
			We examine a choice between bonus contracts offered to dealers of a U.S. auto manufacturer. In our data, dealers select the non-profit-maximizing option in \getval{mistake_rate_pct} percent of observations, costing the mistaken dealers \$\getval{forgone_mean_mistake} per year on average. We examine how the propensity to make this mistake varies with competition, identified both cross-sectionally and within dealers over time. Both analyses show that greater competition substantially lowers the rate of mistakes. However, even in the most competitive markets, consequential mistakes persist. Our results suggest that competition disciplines mainly through within-dealer changes in behavior rather than entry and exit.
		\end{singlespace}
	\end{abstract}
	
	\clearpage
	\section{Introduction} \label{sec:intro}

A burgeoning literature in behavioral economics documents counter-normative decisions across many domains. Despite this evidence, economists commonly argue that such decisions may be safely ignored in many settings because market competition is expected to discipline anomalous decision-making \citep[for classic examples, see][]{alchian1950, friedman1953, becker1962}. \citet{list2003} famously tested this proposition, showing that professional traders of sports memorabilia and collector pins harbored negligible endowment effects, whereas ordinary customers exhibited them.\footnote{Concordant and supporting results also appeared in \cite{list2002}, \cite{list2004}, and \cite{list2004jpe}.} In the years since, this setting has served as the default illustration of the often-made claim that exposure to market competition disciplines anomalies.\footnote{For more recent related results, see \cite{Feng2005}, \cite{Dhar2006}, \cite{Braga2009}, \cite{engelmann2010reconsidering}, \cite{list2011}, \cite{pope2011}, \cite{Apicella2014}, or \cite{Anagol2018}. \label{fn:lit}}

In this paper we present a new test of the relationship between market competition and anomalous decision-making, with anomalies defined as non-profit-maximizing choices. Our setting offers a large and economically important market and financially quite consequential anomalies. Our setting additionally allows us to apply a different approach than \cite{list2003} and subsequent literature by examining variation in the degree of local competition rather than variation in the length of exposure to some fixed market. Like prior literature, we find that competition reduces the prevalence of anomalies. However, even in the most competitive environments we observe, anomalies remain prevalent enough to be consequential. We also document how competition disciplines anomalies: in our setting, it operates primarily by prompting firms to improve their decisions, rather than by driving anomalous ones out of the market. 

We study the dealerships of a large automotive manufacturer operating throughout the U.S. The dealers operate independently, motivating the manufacturer to address principal-agent dynamics through bonus contracts that reward monthly sales. During our three-year study period (2017--2019), the manufacturer offered two contracts that dealers could choose. One was intended to be selected by high-sales dealers, the other by low-sales dealers. On average, these contracts paid dealers \$\getval{bonus_mean_annual} per year and formed an important part of their total earnings. 

This setting is ideal for revisiting the relationship between anomalies and competition for three reasons. First, it contains focal, consequential, and easy-to-identify ``mistakes''---dealers choosing a contract that is not profit-maximizing. Second, it contains variation in competitive environments. One source is cross-sectional: some dealers operate in tightly competitive markets, others in local monopolies. A second source of variation occurs within-dealer-across-time, with some dealers facing local entry, exit, consolidation, or fluctuation in the size of competitors. Third, following dealers over time reveals how they respond to competition, whether by exiting or by changing their decision-making. 

In our setting, dealers frequently engage in anomalous behavior. We find that in \getval{mistake_rate_pct} percent of dealer-years, dealers chose a contract that, ex post, yielded lower profits than the alternative. These mistakes result, on average, in \$\getval{forgone_mean_mistake} of forgone profits per year, and \$\getval{forgone_mean_p90} per year in the top decile. While some of these mistakes reflect uncertainty about future sales, many are predictable from past sales.  

The propensity to make this mistake is strongly related to market competition. In analysis that isolates the cross-sectional relationship, we find that a one-standard-deviation increase in competition lowers the rate of mistakes by \getval{xsec_effect_lo_pct} to \getval{xsec_effect_hi_pct} percent relative to the mean rate, depending on the measure of competition. Comparing the most and least competitive Designated Market Areas (DMAs)---the local markets in which dealers compete---we estimate the difference in the rate of mistakes to be \getval{dma_gap_pct} percentage points (pp). We find similar effects when we exploit within-dealer changes in competition over time. Across our specifications, a one-standard-deviation increase in competition lowers the rate of mistakes by \getval{within_effect_lo} to \getval{within_effect_hi} pp. 

These effects arise through a different channel than the standard account of how competition disciplines a market. Traditionally, the discipline in a competitive market is thought to come from undisciplined agents being forced out of the market by undershooting the zero profit condition. In our setting, exit has a comparatively modest role (perhaps to be expected given our three-year study window). Restricting attention to dealers who remain in the sample for the full study window---a sample in which exit has no role---leaves our estimates of the effect of competition largely unchanged. In contrast to the view in which exit itself is the disciplining factor, these results suggest that competition benefits firms largely by prompting closer attention to their business practices.

Taken together, these results support the claim that competition helps discipline mistakes. However, even in the most competitive markets, we still observe mistakes. The most competitive quartile of dealer-years has a mistake rate of \getval{mistake_rate_q4_pct} percent, for an average forgone value of \$\getval{forgone_mean_q4}. Comparing a perfectly competitive market to a monopoly market, our model suggests mistake rates of \getval{pred_rate_perfcomp_pct} percent and \getval{pred_rate_monopoly_pct} percent, respectively. While competition disciplines, it does so incompletely. Imperfect decisions warrant attention even in a market as ``real'' and competitive as the most competitive segments of the U.S. automotive industry.

We contribute to the literature that documents behavioral anomalies in the field and asks how market structure shapes them. Such anomalies appear across a wide range of field settings \citep{dellavigna2009}. Several market features modulate them. One is experience: repeated trading attenuates the endowment effect among sportscard and pin dealers \citep{list2002, list2003, list2004, list2004jpe}, though it leaves myopic loss aversion intact among professional traders \citep{haigh2005}. A second is the stakes: larger payoffs are thought to sharpen decisions, yet loss aversion survives among professional golfers despite substantial prize money \citep{pope2011}, and game-show contestants make reference-dependent choices even when playing for life-changing sums \citep{post2008}. A third, central to economic theory but far less studied empirically, is competition: markets are thought to discipline non-maximizing firms both by selecting them out \citep{alchian1950, friedman1953} and by pressuring survivors to improve \citep{hart1983, nickell1996, schmidt1997}. Yet direct field evidence on how competition modulates anomalies---and on which of these two channels operates---remains scarce. We study competition as such a feature, among firms facing a continuum of competitive intensity, and distinguish whether it disciplines anomalies through selection or improved decisions.

Our finding contributes to a growing literature on ``behavioral firms'' \citep[for a review, see][]{heidhues_behavioral_2018}. Individual-level biases are well documented; far less is known about their firm-level counterparts, though the evidence is accumulating. Entrepreneurs are overly optimistic about growth \citep{landier2008financial}; managers hold overconfident beliefs about their own performance \citep{huffman2022persistent}; firms treat transitory shocks as permanent \citep{goldfarb2019transitory}; auto dealers act on loss aversion \citep{pierce2020negative}; retailers and ride-hailing platforms forgo profits in the face of consumers' left-digit bias \citep{strulov2019more, list2023leftdigit}; employers set wages coarsely at round numbers \citep{dube2025monopsony, reyes2026coarse}; and managers with weaker cognitive skills adopt mistaken mental models of pricing that persist despite experience \citep{han2026minds}. A related strand documents outright departures from profit maximization \citep[e.g.,][]{hanna2014learning, almunia2021strategic}. We show how firms' anomalous decisions vary with competition and that they persist even where market discipline is strongest.

This paper also contributes to the literature on behavioral contract theory \citep{koszegi2014}. We show that while firms generally self-select into optimal incentive contracts, they often make mistakes, especially absent competitive pressure. The findings align with prior research suggesting that strategic decisions by firms are best explained by models that incorporate both rational intent and behavioral influences \citep{distefano2015, malmendier2015, ellison2018}. The results are also consistent with theories of managerial attention \citep{ocasio1997, dessein2016rational, dessein2021managerial}, in which reduced competition lowers attentional focus, producing suboptimal decisions and weaker performance.

The rest of the paper proceeds as follows. Section~\ref{sec:data} describes our institutional context and data. Section~\ref{sec:contractchoices} presents evidence of suboptimal contract selection. Section~\ref{sec:competition} documents the relationship between suboptimal contract selection and competition. Section~\ref{sec:discussion} concludes. 
	
\section{Institutional Context and Data} \label{sec:data}

\subsection{Setting: The U.S. Automobile Industry}

The U.S. automobile industry operates under a strict regulatory framework that shapes manufacturer-dealer relationships. State franchise laws prohibit manufacturers from selling directly to consumers, requiring them to distribute vehicles through independent dealerships. This institutional arrangement limits manufacturers' control over retail pricing: while manufacturers sell vehicles to dealers at fixed invoice prices, dealers independently determine consumer prices.

Franchise agreements and state laws require manufacturers to treat dealerships uniformly, preventing them from offering differentiated incentive schemes. Yet dealerships exhibit substantial heterogeneity in their organizational structure---ranging from publicly-traded corporations and private equity-owned entities to family businesses and independent operations. This combination of regulatory constraints and dealer heterogeneity limits manufacturers' ability to tailor management strategies to individual dealer characteristics.

To influence dealer behavior within these constraints, manufacturers design incentive programs that reward sales volume. These programs address a misalignment: dealers set a mark-up that trades off per-unit profit against volume, while manufacturers prefer lower mark-ups and higher volume. Volume-based incentive programs have thus become central to manufacturers' distribution strategies and often directly influence executive compensation.

These incentive programs are economically consequential: bonuses from manufacturer programs typically account for 15 to 20 percent of total dealership gross profits, and many new-car departments operate at a loss before bonus payments \citep{haig2023}. By motivating dealers to cut prices, these bonuses also intensify competition and draw more dealerships into the program. Understanding how dealerships select among bonus contracts---and how competitive dynamics shape these choices---is therefore essential for analyzing manufacturer-dealer relationships.

\subsection{The Manufacturer Incentive Program}
\label{sec:program}

We study a major vehicle manufacturer that operates an incentive program across its national dealership network. At the beginning of each year, the manufacturer offers dealers a choice between two incentive contracts (referred to as ``tiers''). Each contract specifies monthly sales targets and corresponding per-vehicle bonuses for eligible models.\footnote{Bonus eligibility requires meeting supplementary requirements, such as completing training programs and participating in a reputation management program. In practice, virtually all dealerships that achieve sales targets satisfy these requirements (Table~\ref{tab:summ}).} Dealers can earn additional bonuses for sales of specific vehicle lines.

The contracts vary across our sample period and differ along two dimensions: bonus rates and enrollment fees. Both contracts feature identical sales targets structured as dealer-specific monthly thresholds. However, the high-bonus contract (henceforth Tier~1) offers larger per-unit bonuses but charges an annual enrollment fee of \$40,000, while the low-bonus contract (henceforth Tier~2) provides smaller bonuses with a \$20,000 annual fee. Additionally, in some years the contract is split into two separate schedules, one for general sales and one for sales of a specific popular vehicle.\footnote{Additionally, across our sample period, several short-term changes to these payment schemes occurred, sometimes mandatory and sometimes at the dealer's discretion. In essentially all cases, these changes increased the returns to selling a particular model (for marketing purposes or to clear excess inventory). In remaining cases, the contract was briefly changed to provide a period of extra motivation for sales.}

\begin{figure}[H]
\caption{The Incentive Contract and Tier Choice} \label{fig:example_contract}
	\centering
	\begin{subfigure}[t]{.48\textwidth}
		\caption*{Panel A. Bonus per Vehicle by Sales Attainment}
		\centering
		\includegraphics[width=\linewidth]{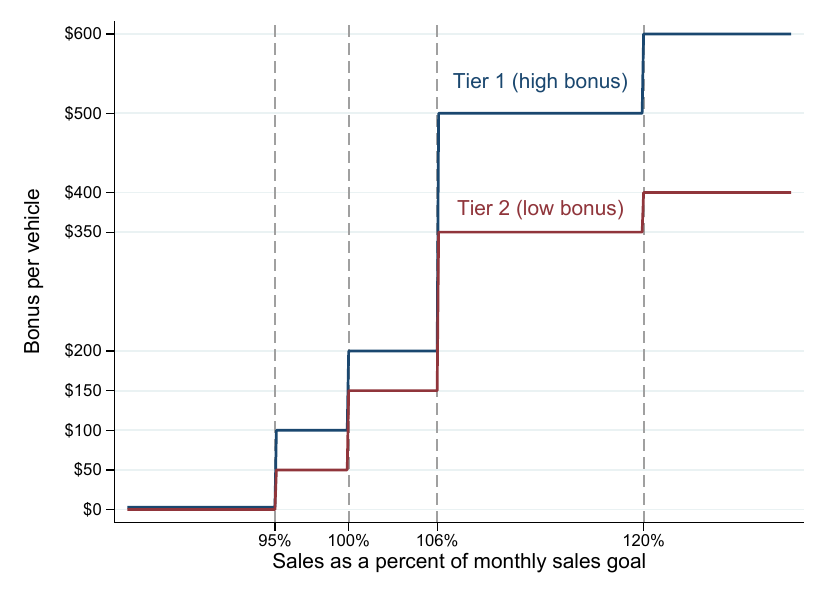}
	\end{subfigure}\hfill
	\begin{subfigure}[t]{0.48\textwidth}
		\caption*{Panel B. Annual Profit for a Median-Size Dealer}
		\centering
		\includegraphics[width=\linewidth]{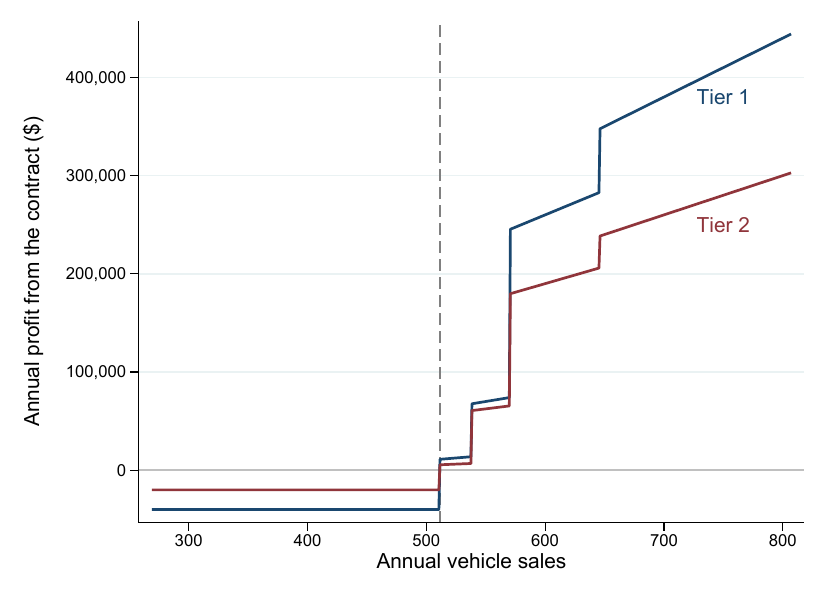}
	\end{subfigure}
	\begin{singlespace}
		\noindent \justify \footnotesize
		\textit{Notes:} This figure illustrates the two incentive contracts using the 2017 schedule. Panel A plots the bonus paid per vehicle as a function of sales attainment, measured as the percent of the dealer's monthly sales goal (its sales in the same month of the previous year), under the high-bonus tier (Tier~1) and the low-bonus tier (Tier~2). Both schedules are step functions with jumps at 95, 100, 106, and 120 percent of the goal. Panel B plots the implied annual profit from each tier, equal to total bonus payments minus the annual enrollment fee (\$40,000 for Tier~1 and \$20,000 for Tier~2), for a median-size enrolled dealer (annual sales of \getval{median_annual_sales} vehicles) as a function of realized annual sales, assuming sales are spread evenly across the year.
	\end{singlespace}
\end{figure}

Figure~\ref{fig:example_contract} illustrates a representative contract from 2017, with discontinuities at four points defined by dealers' ``sales goal'' (their sales that month in the prior year). Tier~1 dealers earned no per-vehicle bonus if they did not reach 95 percent of their sales goal, \$\getval{sched_t1_b2} per sale if they sold between 95 percent and 100 percent of their sales goal, \$\getval{sched_t1_b3} per sale if they sold between 100 percent and 106 percent of their sales goal, \$\getval{sched_t1_b4} per sale if they sold between 106 percent and 120 percent of their sales goal, and \$\getval{sched_t1_b5} if they sold more than 120 percent of their sales goal. Notably, the changing bonus amount is applied to all sales, not just the marginal sales within a given range, leading to discrete jumps in the total bonus when a threshold is passed and the bonus amount is increased. Tier~2 followed the same structure with smaller per-sale bonuses---\$\getval{sched_t2_b2}, \$\getval{sched_t2_b3}, \$\getval{sched_t2_b4}, and \$\getval{sched_t2_b5} across the same bands (about two-thirds of the Tier~1 rates on average). Panel A plots this schedule; Panel B, the annual profit it implies for a median-size dealer under each tier. 

The structure of the 2018 and 2019 contracts are largely similar, with a set of points defined relative to sales goals where per-vehicle bonuses jumps---but with an important difference. For both the 2018 and 2019 contracts, both bonuses and costs are doubled in the high-bonus tier contract. As a consequence, if net profits of low-bonus tier are positive, then net-profits of the high-bonus tier would be double them and larger. If net profits of the low-bonus tier are negative, then they would be for the high-bonus tier as well. With this structure, a risk-neutral firm should not choose the low-bonus tier contract in these years. However, a non-risk-neutral firm could still choose this contract due to its trade off of lower expected payoff for lower variance and lower downside risk.

\subsection{Data and Sample}

Our primary dataset comprises the universe of vehicle sales by U.S. dealerships for a major manufacturer during 2017--2019. The data include both vehicle-level and dealership-level information. Individual vehicles are identified by their 17-character Vehicle Identification Number (VIN), which encodes model characteristics including line, series, and year. Dealerships are identified by unique six-digit Business Associate Codes (BAC), with associated information on location and sales personnel. We aggregate these data to construct a dataset at the dealership-by-month level.

We draw on two manufacturer sources. The quarterly payout reports tell us, for each dealership, the contract it chose, its sales targets, the bonus payments it received, and whether it met the program requirements. Process manuals and dealer communications provide the monthly payout schedules, which we use to recover how each contract's bonus rates map to sales.

Our analysis sample comprises incumbent dealerships---those already operating at the start of our sample period in January 2017---yielding \getval{n_dealers} unique dealerships. This restriction excludes dealerships founded after our sample begins. The resulting dataset contains \getval{n_dealer_months} dealership-month observations spanning 2017--2019.

\subsection{Dealership Characteristics and Contract Choices}

Table~\ref{tab:summ} presents descriptive statistics for our dealership sample. The average dealership employs \getval{sumstat_employees} salespeople and sells \getval{sumstat_sales} vehicles per month, with trucks comprising \getval{sumstat_truck_pct} percent of sales volume (Panel A). Take-up is high: \getval{sumstat_enroll_pct} percent of dealerships enroll for at least one year, with an average enrollment duration of \getval{sumstat_years_enrolled} years. Enrolled dealers are larger in both employment and sales, selling \getval{sumstat_sales_enrolled} vehicles per month (Panel B). Among participants, \getval{sumstat_tier1_pct} percent select the high-bonus Tier~1 contract. 

Most dealerships (\getval{sumstat_same_enroll_pct} percent) maintain their enrollment status across consecutive years---either remaining enrolled (\getval{sumstat_stay_enrolled_pct} percent) or non-enrolled (\getval{sumstat_stay_unenrolled_pct} percent). Among the \getval{sumstat_change_enroll_pct} percent who change enrollment status, switches from non-enrollment to enrollment are more common than exits (\getval{sumstat_enter_enroll_pct} versus \getval{sumstat_exit_enroll_pct} percent). Conditional on enrollment, tier choices show even greater stability: \getval{sumstat_same_tier_pct} percent of enrolled dealers maintain their tier selection across years (Panel C).
	
	\section{Contract Choice by Car Dealerships} \label{sec:contractchoices}

This section examines dealerships' contract enrollment decisions and evaluates their optimality. These contracts are designed to induce dealers to sort based on their expected sales volume, with larger dealerships selecting higher-tier contracts. We first document these sorting patterns empirically, showing how dealership size predicts program participation and tier selection. We then assess whether dealers' choices successfully maximize profits. 

\subsection{Selection into Incentives Contract} 

\begin{figure}[H]
\caption{Enrollment in Incentive Contracts and Dealership Size} \label{fig:enroll}
	\centering
	\begin{subfigure}[t]{.48\textwidth}
		\caption*{Panel A. Distribution of Sales by \\ Enrollment Status}
		\centering
		\includegraphics[width=\linewidth]{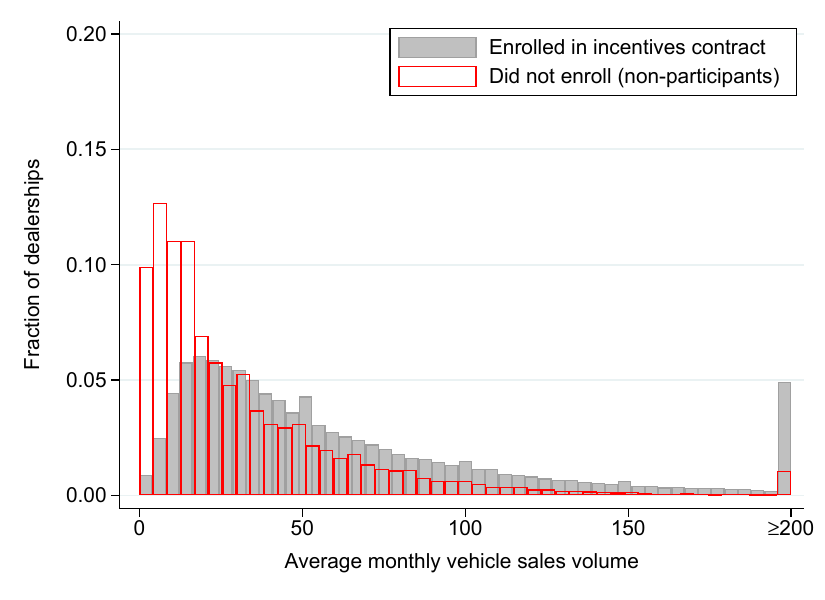}
	\end{subfigure}
	\hfill		
	\begin{subfigure}[t]{0.48\textwidth}
		\caption*{Panel B. Distribution of Sales by \\ Tier Choice}
		\centering
		\includegraphics[width=\linewidth]{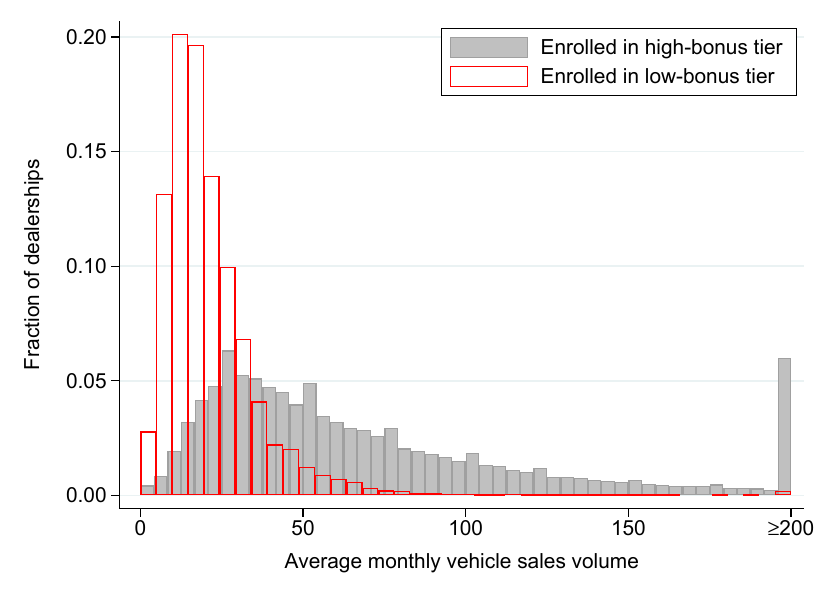}
	\end{subfigure}	
	\hfill		
	\begin{subfigure}[t]{.48\textwidth}
		\caption*{Panel C. Program Enrollment and \\ Monthly Sales}
		\centering
		\includegraphics[width=\linewidth]{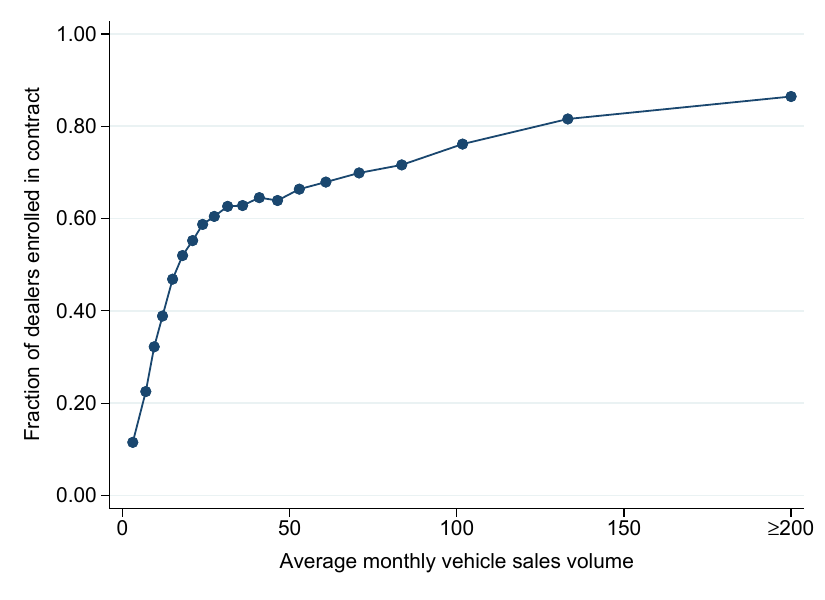}
	\end{subfigure}
	\hfill		
	\begin{subfigure}[t]{0.48\textwidth}
		\caption*{Panel D. Tier Selection and \\ Monthly Sales}
		\centering
		\includegraphics[width=\linewidth]{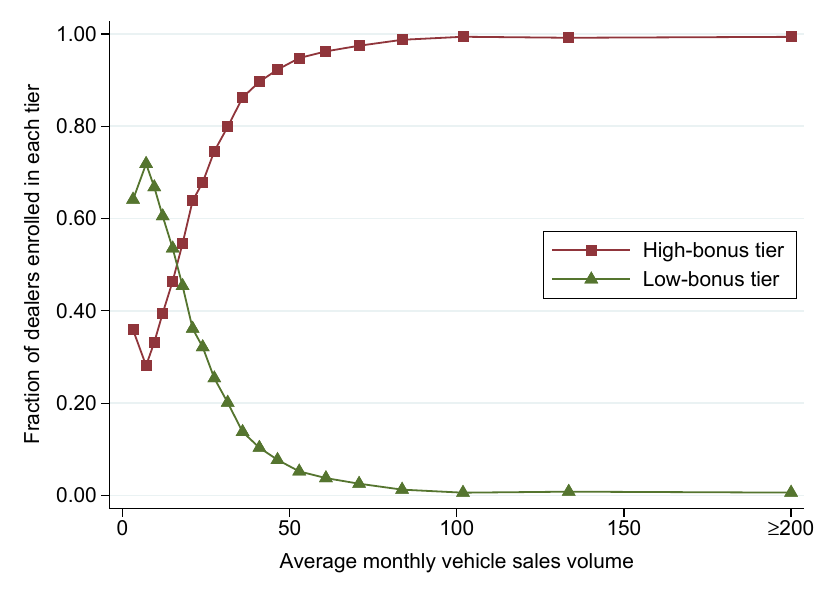}
	\end{subfigure}
	{\footnotesize
		\singlespacing \justify
		
		\textit{Notes:} This figure shows the relationship between dealership size and incentive program participation. Panels A and B show distributions of average monthly vehicle sales for dealerships by enrollment status and tier choice, respectively. Panels C and D present binned scatterplots (20 equal-sized bins) of enrollment rates against monthly sales volume. In Panel C, the $y$-axis shows the fraction of dealerships enrolled in either tier. In Panel D, separate lines indicate enrollment rates for the high-bonus tier (maroon) and the low-bonus tier (green).
					
	}
\end{figure}

Dealership size strongly predicts program participation and tier selection, though substantial heterogeneity remains among similarly sized dealers. Figure~\ref{fig:enroll} documents these patterns by examining the relationship between dealership size (monthly sales volume) and contract choices.

Panels A and B display the distribution of monthly vehicle sales by enrollment status and tier choice. Program participants (gray bars) exhibit higher sales volumes than non-participants (red bars), though the distributions overlap considerably (Panel A). Among enrolled dealerships (Panel B), tier selection correlates strongly with size: low-bonus tier dealers (red bars) concentrate below 50 vehicles per month, while high-bonus tier dealers (gray bars) typically operate at higher volumes. Yet sales volume does not perfectly predict tier choices. The two sales distributions have substantial overlap in the 20--40 vehicle range, showing that some dealers facing similar demand patterns make different tier choices.

Panels C and D present binned scatterplots that help quantify these relationships. Panel C reveals that enrollment probability increases roughly monotonically with dealership size. The relationship is steepest for dealerships selling fewer than 50 vehicles monthly, where enrollment rates rise from 10 percent to over 60 percent. For larger dealerships, enrollment continues to increase but at a diminishing rate, approaching---though never reaching---universal participation. Conditional on enrollment, dealerships selling fewer than 20 vehicles monthly predominantly select the low-bonus tier (Panel D). Between 20 and 50 vehicles, the probability of choosing the high-bonus tier increases sharply, creating a transition region where both tier choices are common. Beyond 50 vehicles monthly, virtually all enrolled dealerships opt for the high-bonus tier.

These patterns reveal a key feature of contract selection. Size strongly predicts participation and tier choice, but the imperfect sorting suggests that other factors---such as local market conditions or managerial ability---may also shape these decisions.\footnote{\cite{list2003} and much of the subsequent literature (see footnote \ref{fn:lit}) find that experienced actors show fewer anomalies. To assess this claim, we compare the suboptimal choices of experienced firms---those present before our sample period---with those of firms created during our sample period (and thus not part of our main sample). New entrants make ex-post suboptimal choices at a rate of \getval{mistake_new_pct} percent, compared to \getval{mistake_exp_pct} percent for experienced dealerships---a gap of \getval{mistake_exp_gap_pp} pp that represents a \getval{mistake_exp_relinc_pct} percent increase in mistake rates. This pattern is consistent with experienced dealerships having higher-quality decision-making, whether through selection or learning.}

\subsection{Ex-Post Non-Profit-Maximizing Contract Choices} \label{sub:mistakes}

We assess the ex-post optimality of dealerships' contract choices by comparing actual profits with counterfactual profits under alternative contracts. This analysis proceeds in two steps.

First, we compute counterfactual bonuses that each dealership would have received under alternative contract choices (Tier~1, Tier~2, or nonparticipation). This analysis is restricted to dealerships that participated in the incentive program at least once during our sample period, as we cannot observe the sales targets that never-participants would have faced. For dealership $i$ that chose contract $j$, we calculate what their bonus payments would have been under contracts $j' \neq j$ using their observed sales. This calculation assumes no behavioral response; that is, sales would remain at the same level regardless of contract choice. Because a more generous contract raises the marginal return to effort, allowing sales to respond would only widen the bonus gap in favor of the better contract. Our estimates should thus be interpreted as a conservative lower bound on the fraction of suboptimal choices.\footnote{One caveat to this interpretation is that our analysis does not account for any costs incurred to comply with program requirements. In practice, these costs are real but small. Accounting for these costs could rationalize some non-participants who forgo small profits from participation, but cannot rationalize forgoing more substantial profits or choosing the wrong tier conditional on program participation.}

For each contract option $j \in \{0, 1, 2\}$---non-participation, Tier~1, and Tier~2, respectively---we compute annual profits as:
\begin{align*}
	\pi_{i,j} = \sum_{t=1}^{12} \text{Bonus}_{i,j,t} - k_j,
\end{align*}
where $\text{Bonus}_{i,j,t}$ represents monthly bonus payments and $k_j$ is the annual enrollment fee. We classify a dealership's choice as ex-post suboptimal if an alternative contract would have yielded higher profits given realized sales.

Table~\ref{tab:ctfl} reports the fraction of dealerships making suboptimal choices (Panel A) and decomposes these mistakes by the profit-maximizing alternative (Panel B).

In \getval{mistake_rate_pct} percent of participating dealer-years, the dealership selected a contract that failed to maximize ex-post profits (column 1, $p < 0.01$), a substantial deviation from profit maximization. These mistakes break down by the tier chosen: \getval{mistake_low_pct} percent of participants chose the low-bonus tier when it was suboptimal (column 3), compared to only \getval{mistake_high_pct} percent who chose the high-bonus tier when it was suboptimal (column 2, both $p < 0.01$). The \getval{mistake_tier_gap_pp} pp difference is statistically significant ($p < 0.01$, with standard errors clustered by dealership).

The predominant error involves under-investment. Among all participating dealerships, \getval{mistake_should_high_pct} percent would have earned higher profits by switching to the high-bonus tier, whereas only \getval{mistake_should_low_pct} percent would have earned higher profits by switching to the low-bonus tier (Panel B, column 1).\footnote{This asymmetry is to be expected since switching to the low-bonus tier cannot be ex post optimal in 2018 and 2019, as discussed in Section \ref{sec:program}.} A further \getval{mistake_program_pct} percent of participants would have maximized profits by forgoing the program entirely; this mistake is somewhat more common among low-bonus tier enrollees (\getval{mistake_program_low_pct} percent, column 3) than high-bonus tier enrollees (\getval{mistake_program_high_pct} percent, column 2).

These patterns persist across individual years (Appendix Tables~\ref{tab:ctfl_2017}--\ref{tab:ctfl_2019}), suggesting that suboptimal choices reflect systematic decision-making failures rather than unusual market conditions in any particular year. 

\subsection{Quantifying the Cost of Suboptimal Choices} \label{sub:cost}

How costly are these ex-post non-profit-maximizing choices? To answer this, we compute the forgone profits from these decisions. For each dealership $i$ that chose contract $j \neq 0$, we measure the gain along each margin of the decision:
\begin{align*}
	\text{Gain from participation}_{i} &= \pi_{i,j} - \pi_{i,0} = \sum_{t=1}^{12} \text{Bonus}_{i,j,t} - k_j, \\
	\text{Gain from tier choice}_{i} &= \pi_{i,j} - \pi_{i,j'\neq j} .
\end{align*}

The first is the gain from enrolling rather than staying out; it is negative---a loss bounded by the enrollment fee---when non-participation would have been optimal. The second is the gain from the chosen tier relative to the alternative; it is negative when the other tier would have been more profitable. A suboptimal choice thus registers as a negative gain, and its cost is the magnitude of that shortfall.

\begin{figure}[H]
	\caption{Distribution of Realized Versus Counterfactual Profits} \label{fig:hist_profit}
	\centering
	\begin{subfigure}[t]{.48\textwidth}
		\caption*{Panel A. Profit Gain from Program Enrollment}
		\centering
		\includegraphics[width=\linewidth]{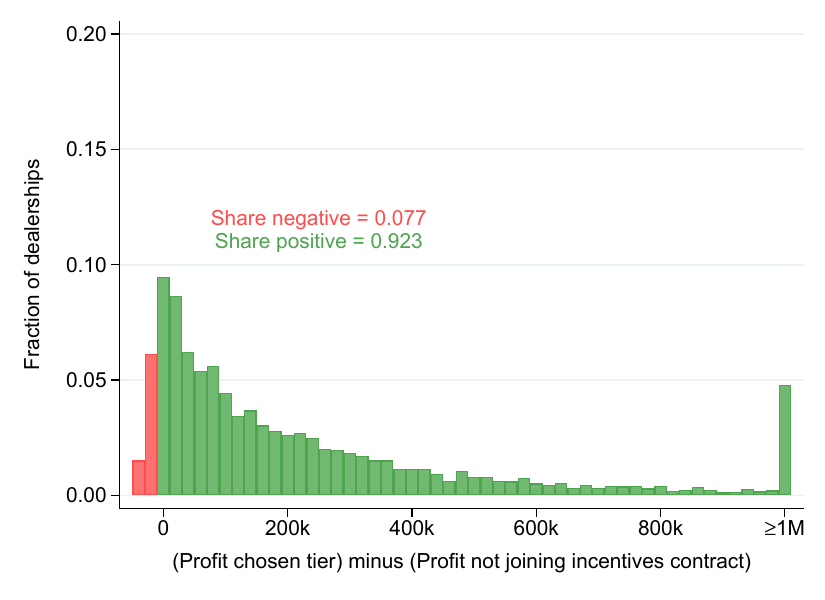}
	\end{subfigure}
	\hfill		
	\begin{subfigure}[t]{0.48\textwidth}
		\caption*{Panel B. Profit Difference Between Chosen and Alternative Tier}
		\centering
		\includegraphics[width=\linewidth]{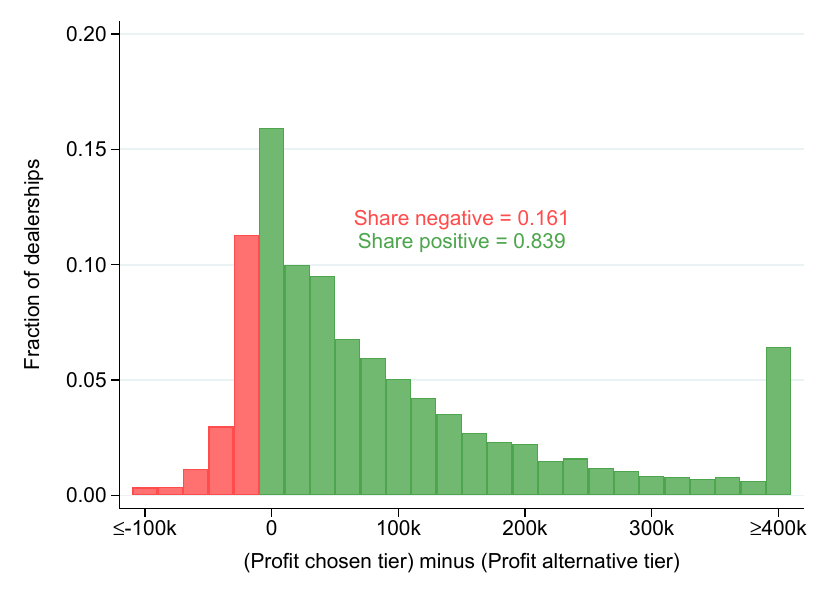}
	\end{subfigure}	
	
	{\footnotesize
		\singlespacing \justify
		
		\textit{Notes:} This figure shows the distribution of profit differences under alternative contract choices. Panel A shows the distribution of profit gains from enrolling in the chosen tier relative to non-enrollment. Panel B displays profit differences between the chosen tier and the alternative tier. Green bars indicate positive profit differences (optimal choices); red bars indicate negative differences (suboptimal choices). Values are calculated using observed sales with no behavioral adjustment. The text in each panel reports the share of dealerships with negative and positive profit differences.
		
	}
\end{figure}

Figure~\ref{fig:hist_profit} presents the distribution of these two variables. Program participation was profitable for most dealerships: only \getval{shr_noenroll_better_pct} percent would have earned higher profits by not enrolling (Panel A). For these dealerships, losses are bounded by the enrollment fee---up to \$20,000 for Tier~2 enrollees and \$40,000 for Tier~1 enrollees---reaching the cap when a dealership pays the fee but earns no bonus in any of the 12 months.

The costs of selecting the wrong tier are substantially larger (Panel B). While \getval{shr_wrongtier_pct} percent of participating dealerships chose a suboptimal tier, the financial consequences vary widely. The distribution exhibits a long left tail: most dealerships making suboptimal tier choices forgo less than \$10,000, but some sacrifice over \$100,000 in potential profits. These large losses occur when dealerships select the low-bonus tier despite having sales volumes that would trigger substantial bonuses under the high-bonus tier.

Overall, the average cost of committing \emph{any} mistake (choosing the wrong tier or enrolling when it was not worthwhile) is \$\getval{forgone_mean_mistake} per year. Among participants for whom the other tier would have been more profitable, the average profit forgone by not switching tiers is \$\getval{cost_wrongtier} per year.

\begin{figure}[H]
	\caption{Contract Choice and Financial Stakes}\label{fig:choice_tier1_profit}
	\centering
	\includegraphics[width=.75\linewidth]{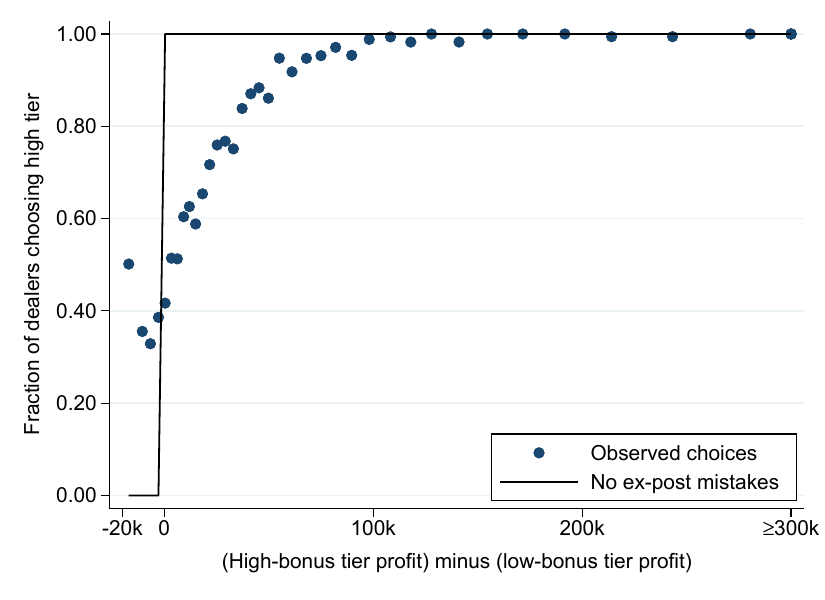}
	
	{\footnotesize
		\singlespacing \justify
		
		\textit{Notes:} This figure shows the relationship between tier selection and profit differential. The figure displays a binned scatterplot (40 equal-sized bins) of high-bonus tier enrollment rates against the ex-post profit difference between the high-bonus and low-bonus tiers. Profit differences are calculated using observed sales with fixed tier-specific bonus structures and enrollment fees, assuming no behavioral response (see Section~\ref{sub:mistakes} for details). The solid black line indicates the theoretical optimum where dealers choose the high-bonus tier whenever the profit difference is positive. Deviations from this line represent ex-post suboptimal choices.
		
	}
\end{figure}

Dealers' tier choice responds to the difference in financial returns across tiers. To illustrate, Figure~\ref{fig:choice_tier1_profit} plots the probability of choosing the high-bonus tier as a function of the cross-tier profit differential. Under perfect profit maximization (black line), dealerships would exhibit a step function: choosing the high-bonus tier if and only if it yields higher profits. The empirical relationship (blue circles) shows sensible patterns, but deviates from this benchmark. As the returns to the high-bonus tier grow sufficiently large, the probability of selecting it approaches 1. Concretely, dealers who stand to gain \$100,000 or more by selecting the high-bonus tier rarely fail to do so.\footnote{Appendix Figure~\ref{fig:mistake_size} plots the mistake rate by dealership size. It shows that the likelihood of a mistake becomes small for sufficiently large dealers.} This ``perfection in the tails'' result is not symmetric. In cases where the low-bonus tier is strongly financially incentivized (higher-tier profits - lower-tier profits $\approx-\$20,000$), approximately \getval{choice_high_lowinc_pct} percent of dealerships still select the high-bonus tier. Because the low-bonus tier is designed for smaller dealerships, its benefits cannot grow as large as the high-bonus tier's, generating the asymmetry in the plot. 

While some degree of deviation from the step function is expected, the degree observed in this figure cannot easily be rationalized with sales data. Consider, for example, dealers who chose suboptimally with a high-bonus tier versus low-bonus tier profit differential between -\$10,000 and -\$20,000. These are dealers who selected the high-bonus tier but would be better off having selected the low-bonus tier. The median dealer in this group would have to have predicted that their sales would grow by at least \getval{re_overinvest_growth_pct} percent from the prior year to justify their choice. In practice, sales growth by more than 10 percent occurs in only \getval{growth_gt10_pct} percent of dealer-years, and sales growth of more than 30 percent occurs in only \getval{growth_gt30_pct} percent of dealer-years (see Appendix Figure~\ref{fig:yoy_sales}). Consider next dealers who chose suboptimally with a high-bonus tier vs low-bonus tier profit differential between \$10,000 and \$20,000. These are dealers who selected the low-bonus tier but would be better off having selected the high-bonus tier. The median dealer in this bin would have to have predicted that their sales would shrink by at least \getval{re_underinvest_shrink_pct} percent from the prior year to justify their choice. In practice, sales shrinkage by more than 14 percent occurs in only \getval{shrink_gt14_pct} percent of dealer-years. Results are starker farther up the curve: mistaken dealers with a high-bonus tier vs low-bonus tier profit differential between \$50,000 and \$60,000, on median, must have expected sales shrinkage by more than \getval{re_underinvest_shrink_far_pct} percent, which occurs in only \getval{shrink_gt27_pct} percent of dealer-years. In short, while it is difficult to judge the wisdom of any individual decision because of the potential for private information, in aggregate these decisions are at odds with rational expectations.
	
	\section{Market Competition and Contract Choice} \label{sec:competition}

Having established that a substantial fraction of dealerships make suboptimal contract choices, we now examine whether market competition disciplines this behavior. 

\subsection{Measuring Competition in Local Car Markets}

We measure competition at the DMA level, which corresponds to the local market in which dealerships sell and advertise.\footnote{DMAs are Nielsen-defined local media markets, each a mutually exclusive group of whole counties in which a common set of broadcast stations predominates. Because dealerships advertise and consumers shop within these markets, the DMA is a natural unit for measuring local competition.} For each DMA $j$ and year $t$, we construct four leave-one-out measures of competitive intensity:

\emph{Herfindahl-Hirschman Index (HHI):} We calculate the HHI based on dealerships' shares of total vehicle sales within each DMA:
\begin{align*}
	\text{HHI}_{jt} = \sum_{d \in \mathcal{D}_{jt}} s_{djt}^2, \quad \text{where} \quad s_{djt} = \frac{\text{Sales}_{djt}}{\sum_{d' \in \mathcal{D}_{jt}} \text{Sales}_{d'jt}},
\end{align*}
and $\mathcal{D}_{jt}$ denotes the set of dealerships in DMA $j$ in year $t$. The HHI ranges from near zero (perfect competition) to one (monopoly), with higher values indicating greater market concentration.

\emph{Top-$n$ Market Share:} As a complementary concentration measure, we compute the combined market share of the largest dealerships:
\begin{align*}
	\text{Top-}n\text{ Share}_{jt} = \sum_{q=1}^{n} s_{qjt},
\end{align*}
where $s_{qjt}$ denotes the market share of the top $q$ largest dealership in DMA $j$. We compute this measure for $n \in \{1, 3, 5\}$. 

For every measure we adopt a leave-one-out construction: in computing the concentration faced by a given dealership, we exclude that dealership's own sales and renormalize the remaining shares. This construction avoids endogeneity problems that would arise from a dealer's own sales influencing both the competition measure and whether their contract is optimal.

Table~\ref{tab:summ}, Panel D presents summary statistics for our competition measures. The average DMA has an HHI of \getval{comp_hhi_mean_dma}, indicating relatively low concentration typical of competitive markets. However, concentration varies substantially both across markets (HHI ranges from near zero to one) and within markets over time. Across DMAs, the largest dealership captures on average \getval{comp_top1_mean_pct} percent of the market, while the top 3 dealerships combined control \getval{comp_top3_mean_pct} percent and the top 5 control \getval{comp_top5_mean_pct} percent.

\subsection{Descriptive Evidence}

\begin{figure}[H]
\caption{Market Concentration and Ex-Post Suboptimal Choices}\label{fig:binsc_comp}
	\centering
	\begin{subfigure}[t]{.48\textwidth}
		\caption*{Panel A. Market Share of \\ Largest Dealer}
		\centering
		\includegraphics[width=\textwidth]{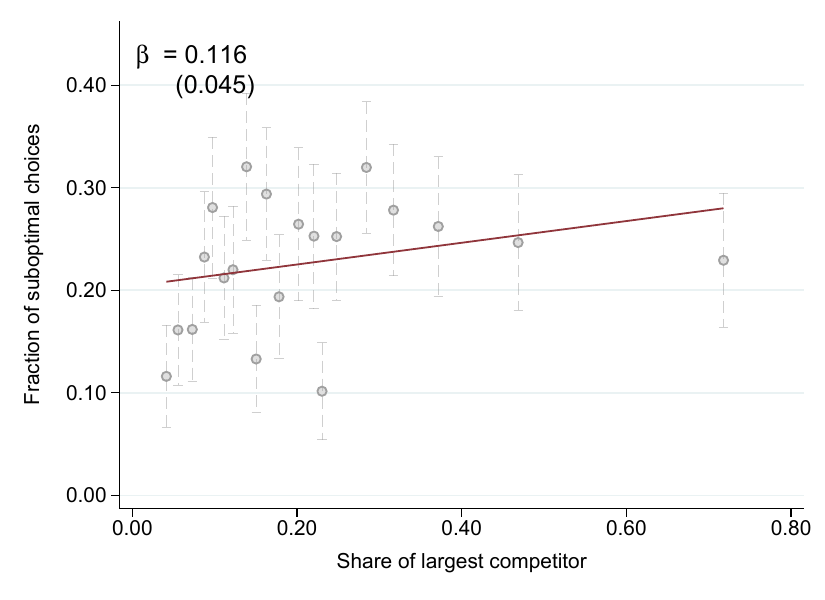}
	\end{subfigure}
	\hfill
	\begin{subfigure}[t]{.48\textwidth}
		\caption*{Panel B. Market Share of \\ Top 3 Dealers}
		\centering
		\includegraphics[width=\textwidth]{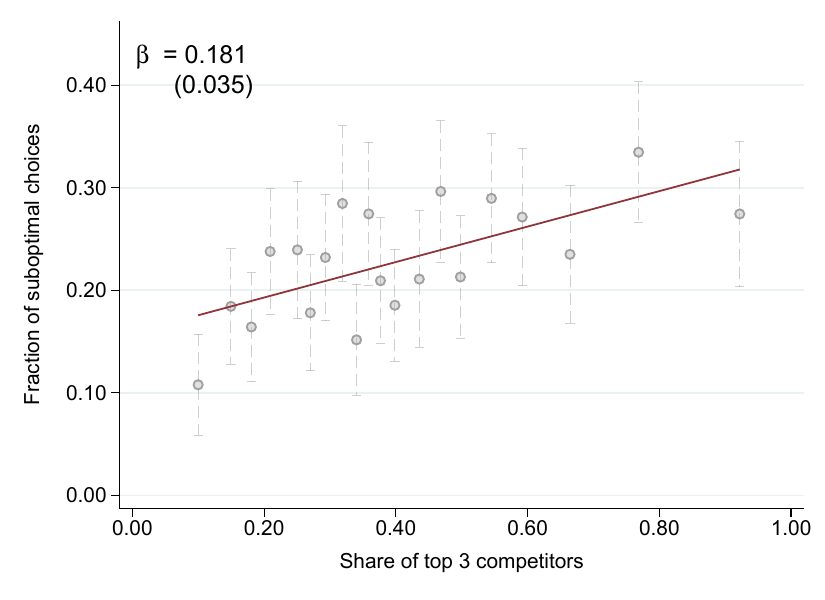}
	\end{subfigure}
	\hfill
	\begin{subfigure}[t]{.48\textwidth}
		\caption*{Panel C. Market Share of \\ Top 5 Dealers}
		\centering
		\includegraphics[width=\textwidth]{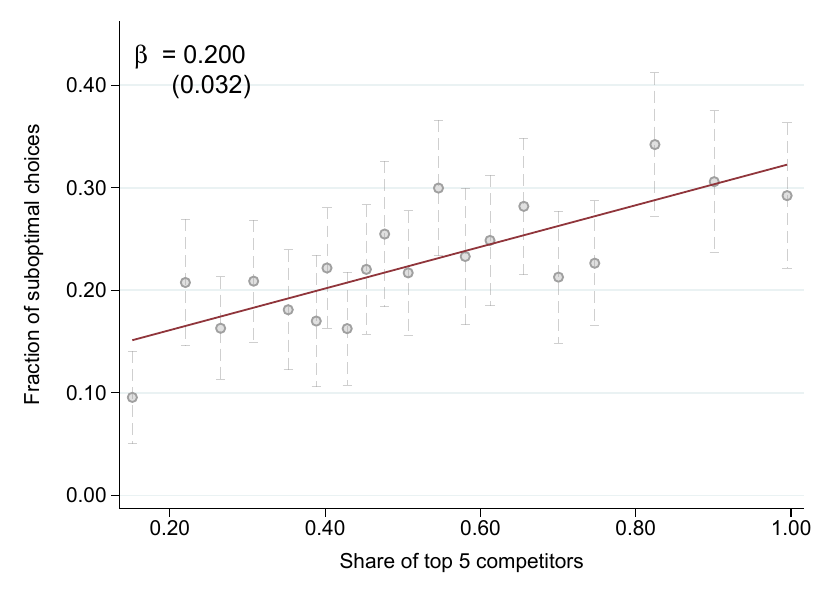}
	\end{subfigure}
	\hfill
	\begin{subfigure}[t]{.48\textwidth}
		\caption*{Panel D. Herfindahl-Hirschman \\ Index (HHI)}
		\centering
		\includegraphics[width=\textwidth]{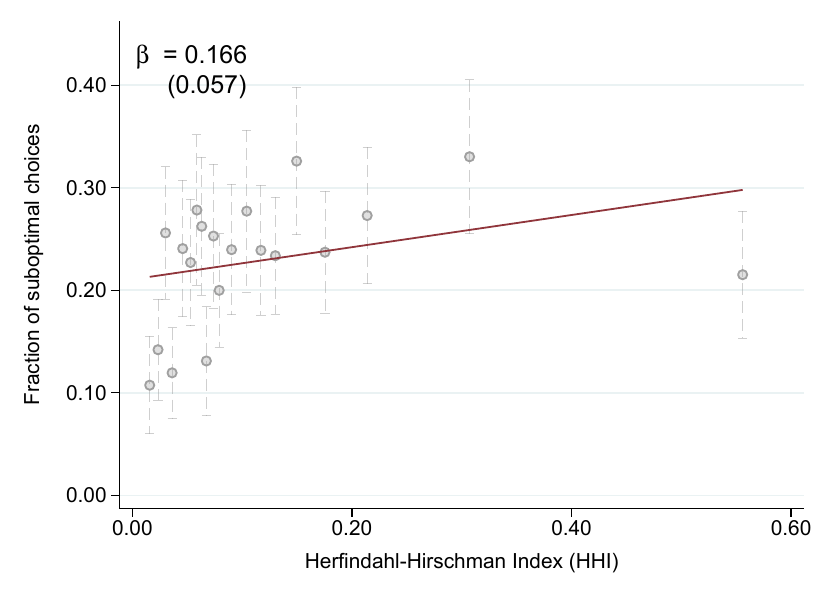}
	\end{subfigure}

	{\footnotesize
		\singlespacing \justify

		\textit{Notes:} This figure shows the relationship between market concentration and ex-post suboptimal contract choices. Binned scatterplots (20 equal-sized bins) with 95 percent confidence intervals show the fraction of dealerships making suboptimal tier choices against each concentration measure. Each dealership enters as a single cross-sectional observation, averaging its suboptimal-choice indicator and its market's concentration across 2017--2019. An ex-post mistake occurs when a dealership's realized profit under its chosen contract is lower than under the best alternative. Panel A uses the market share of the largest dealer in each DMA; Panel B uses the combined share of the top 3 dealers; Panel C uses the combined share of the top 5 dealers; Panel D uses the Herfindahl-Hirschman Index. The maroon line shows the best linear fit.

	}

\end{figure}

Before presenting regression estimates, we provide visual evidence on the relationship between market concentration and contract choice optimality. Figure~\ref{fig:binsc_comp} presents binned scatterplots relating our four concentration measures to the fraction of dealerships making ex-post suboptimal tier choices (see Appendix Figures~\ref{fig:binsc_comp_2017}--\ref{fig:binsc_comp_2019} for the same relationship estimated separately within each year). 

Across all four measures, more concentrated markets have higher rates of suboptimal choices. Moving from the most competitive markets (where the largest dealer controls approximately \getval{binsc_xlo_pct} percent of sales) to the most concentrated markets (where the dominant dealer captures nearly \getval{binsc_xhi_pct} percent) is associated with a \getval{binsc_gap_pp} pp increase in suboptimal choices---from \getval{binsc_lo_pct} percent to \getval{binsc_hi_pct} percent (Panel A). This pattern is consistent across alternative measures of market concentration (Panels B--D).

\subsection{Regression Models}

We use two approaches to estimate the effect of market competition on contract choice optimality. Our first specification includes only time fixed effects: 
\begin{align} \label{eq:timefe}
	\text{Suboptimal}_{it} = \alpha + \beta \cdot \text{Competition}_{j,t} + \lambda_t + \varepsilon_{it},
\end{align}
where $\text{Suboptimal}_{it}$ is an indicator equal to one if dealership $i$ in year $t$ selected a contract tier that yielded lower profits than the best alternative, $\text{Competition}_{j,t}$ represents one of the four measures of competitive intensity in dealership $i$'s DMA $j$ at time $t$, and $\lambda_t$ is a year fixed effect. When fixed effects are included in our regressions they are normalized to have a mean of zero, allowing us to interpret the constant $\alpha$ as the predicted value of the dependent variable for the average year (or, in the next specification, average firm) under perfect competition.  

Our second specification includes both time fixed effects and dealer fixed effects ($\mu_i$): 
\begin{align} \label{eq:twfe}
	\text{Suboptimal}_{it} = \alpha + \beta \cdot \text{Competition}_{j,t} + \mu_{i} + \lambda_t + \varepsilon_{it}
\end{align}

We estimate these equations using OLS with standard errors clustered at the DMA level. In equation~\eqref{eq:timefe}, our coefficient of interest, $\beta$, measures how choices respond to the cross-sectional variation in competition. In equation~\eqref{eq:twfe}, $\beta$ measures how individual dealers' choices respond to within-DMA changes in concentration over time.\footnote{Within-market variation in concentration stems from four sources. First, dealer consolidation through mergers and acquisitions increases concentration as the number of independent competitors decreases and market shares become more concentrated among fewer firms. Second, dealer exits generate variation in concentration, though the direction of the effect depends on the size of the exiting dealer---the closure of a large dealer may decrease concentration by allowing smaller dealers to expand their market shares, while the exit of a small dealer typically increases concentration. Third, the entry of new dealers changes concentration by adding a competitor and dividing sales among more firms. Fourth, differential growth among existing dealers shifts concentration as some gain market share at others' expense.} A positive $\beta$ indicates that less competitive/more concentrated markets have more suboptimal contract choices. 

Our identifying assumption is that, absent changes in market concentration, the probability of making suboptimal contract choices would have evolved similarly across DMAs (i.e., the standard parallel trends assumption). This assumption would be violated if changes in concentration were driven by factors that directly affect decision-making quality---for instance, if acquiring firms systematically differed from target firms in their managerial ability. The assumption would also fail if time-varying local shocks simultaneously affected both market structure and contract choices, such as local economic downturns that both forced dealer exits and impaired decision-making capabilities.

\subsection{Regression Estimates}

Table~\ref{tab:reg_competition} presents our main results on how market competition affects contract choice optimality. Panel A reports estimates of equation~\eqref{eq:timefe} (which employs cross-sectional variation). Panel B reports estimates of equation~\eqref{eq:twfe} (which employs within-dealer-across-time variation). Each column reports estimates using different measures of market concentration. 

We find that market concentration increases suboptimal contract choices. Focusing first on the cross-sectional results in Panel A, our estimates imply that markets with perfect competition make suboptimal choices \getval{yrfe_pp_lo}--\getval{yrfe_pp_hi} pp less often than markets with monopoly, with three of the four estimates significant at least at the 5-percent $\alpha$-level and the fourth (the largest-dealer share) significant at the 10-percent $\alpha$-level. Relative to the constant---interpreted in the model as the suboptimal-choice rate under perfect competition---this implies a \getval{yrfe_rel_lo}--\getval{yrfe_rel_hi} percent increase under monopoly compared to perfect competition.

Focusing next on the within-firm-across-time results in the bottom panel, our results are qualitatively similar but larger in magnitude. Our estimates imply that markets with perfect competition make suboptimal choices \getval{within_pp_lo}--\getval{within_pp_hi} pp less often than markets with monopoly. Two of the four estimates---the largest-dealer and top-5 shares---are significant at the 5-percent level; the top-3 share and HHI estimates are not statistically significant at traditional levels but are not far off ($p = \getval{within_top3_pval}$ and $p = \getval{within_hhi_pval}$). The estimates are robust to dropping outlier markets (those whose concentration lies more than two or three standard deviations from the mean): the coefficients remain stable or grow larger (Appendix Tables~\ref{tab:reg_competition_out2sd} and~\ref{tab:reg_competition_out3sd}).

Taking stock: across simple descriptive analysis as well as regression analyses based on two identification strategies and four measures of competition, we find strong support for a disciplining effect of competition. However, in almost all analyses (excepting Table~\ref{tab:reg_competition} Panel B column 3), we find strong evidence for mistakes remaining even under perfect competition.  This persistence echoes evidence that behavioral biases survive even among experienced professionals facing high stakes and intense competition---loss aversion among elite golfers \citep{pope2011}, and mispricing and myopic loss aversion among financial traders \citep{coval2005, haigh2005}. 

\subsection{Effort versus Selection}

Competition could improve market-level contract selection through two channels, each corresponding to a long-standing view of how competition disciplines firms. The first is a selection channel, reflecting the survival, or evolutionary, view: competition drives out firms that fail to maximize profits, so that the surviving population behaves \emph{as if} optimizing, whatever the sophistication of any individual firm \citep{alchian1950, friedman1953}. The second is an effort channel, reflecting a managerial-effort-and-attention view: competitive pressure leads the firms that remain to devote more effort \citep{hart1983, nickell1996, schmidt1997} and attention \citep{ocasio1997, dessein2016rational, dessein2021managerial} to their decisions, improving the choices of survivors.

Our baseline estimates combine both effects. To isolate the effort channel, Table~\ref{tab:reg_competition_noexit} reproduces Table~\ref{tab:reg_competition} excluding \getval{n_exit_dealers} dealerships (\getval{exit_share_pct} percent of dealerships) that exit the market at any point during our sample period. This restriction shuts down the selection mechanism, so the remaining variation identifies how competition affects decision-making among firms that survive the full sample period.\footnote{This approach likely overstates the effort effect unconditional on survival. Firms that exit the market are plausibly those that could not or would not improve their decision-making in response to competitive pressure. By restricting the analysis to survivors, we estimate the effect only among firms responsive to competition, thereby overstating the average effort effect across all firms.}

In this analysis, evidence that competition disciplines anomalies is stronger than in the unconditional analysis. Estimated coefficients are nearly identical and overall are statistically stronger, with all estimates significant at least at the 10-percent $\alpha$-level. To illustrate, the within-firm estimate generated using the largest competitor's market share is \getval{noexit_top1_coef} (Panel B, column 1), essentially unchanged from the \getval{within_top1_coef} estimate from the full sample under the same specification (Table~\ref{tab:reg_competition}, Panel B). Because shutting down the selection margin leaves the relationship intact---and this holds across all four concentration measures---competition disciplines contract choices by improving decision-making among surviving firms, not by driving poor decision-makers out of the market.

As further evidence that exit does not account for our results, Appendix Table~\ref{tab:exit} reproduces Table~\ref{tab:reg_competition} but uses exit as the dependent variable. All estimates in this table are statistically insignificant and quantitatively small.\footnote{In the cross-sectional analysis, the largest estimate implies that moving from perfect competition to perfect monopoly has a \getval{exit_xsec_pp} pp effect on the rate of exit. In the within-dealer analysis, the largest coefficient implies that moving from perfect competition to perfect monopoly has a \getval{exit_within_pp} pp effect on the rate of exit.} 

Taken together, these analyses illustrate that, in this setting, the disciplining effect of competition does not come from exit itself, but rather from the improved decision-making of firms remaining in the market. This improvement among survivors reflects the effort channel, not selection. This result is consistent with \citeauthor{backus2020}'s (\citeyear{backus2020}) results from the ready-mix concrete industry, which show a productivity response to competition driven  by within-firm channels (such as managerial inputs) rather than through selection.

\subsection{Forecastability as a Confound}

Optimally choosing the contract tier requires forecasting next year's sales at the time of contract selection. If more competitive markets have sales patterns that are easier to forecast, this could partly explain the competition--suboptimality relationship that we document. This could lead to a different interpretation of our results than we have advanced: rather than suggesting that competition directly affects decision-making (\textit{ceteris paribus}), the results would instead show that competition is associated with easier decisions. 

While we find some mixed evidence that this concern could be present in cross-sectional analysis, we find no such relationship once we add dealer fixed effects. Table~\ref{tab:pred_reg} presents regressions using three measures of forecastability: the \textit{forecast error} (squared residual from predicting a dealer's annual sales from its prior-year sales), the \textit{sales variability} around the dealer's own mean $(\text{Sales}_{it}-\overline{\text{Sales}}_i)^2$, and the \textit{relative variability} $((\text{Sales}_{it}-\overline{\text{Sales}}_i)/\overline{\text{Sales}}_i)^2$. We regress each of these dependent variables on each of the four competition measures, adding time and dealer fixed effects across specifications. 

In regressions with concentration alone or time fixed effects, we find some statistically significant results whose sign depends on the measure used. Overall, concentration appears negatively related to forecast error and to absolute variability---suggesting that more-concentrated markets are easier to forecast and less volatile---but the sign flips to positive and significant for relative variability. While these results warrant some caution in interpreting our cross-sectional estimates, none of the 12 regressions that include time and dealer fixed effects show a statistical relationship between forecastability and competition. We view robustness to this potential confound as one of several advantages of our within-dealer-across-time identification strategy.


	\section{Discussion} \label{sec:discussion}

In this paper we have revisited a classic question: do competitive markets discipline economically ``anomalous'' behaviors? Our answer accords with the classic results of \citet{list2002, list2003, list2004, list2004jpe} in suggesting that they do. However, our setting has several advantages. Compared to prior studies, our primary advantages are (i) a substantially larger and more economically important market, (ii) counter-normative decisions of larger financial consequence, and (iii) actors facing a continuum of competitive intensity rather than a contrast between inexperienced individuals and individuals with some degree of market exposure. These advantages give our findings a stronger basis for addressing this often-posed question. 

We also find two new nuances. First, despite competition's large and positive effect on decision quality, a substantial amount of suboptimal behavior remains even in the most competitive markets we observe. This suggests that competition, while helpful, does not rule out a role for mistakes or behavioral-economic forces. This matters especially because the actors in our setting are firms, not individuals, amplifying the need for research on ``behavioral firms'' (see \citet{heidhues_behavioral_2018} or the papers we discuss in the introduction). Second, competition's positive effects arise from leading firms to improve their decision-making rather than from selecting for good decision-makers. This result supports prior claims that reduced competition promotes suboptimal decisions through reducing managerial attention (see, e.g., \citet{ocasio1997}).

Our results also speak to the principal-agent literature on screening and self-selection. The manufacturer's two-tier menu is a screening device, designed so that each dealer self-selects into the tier suited to its sales volume \citep{bolton2004, koszegi2014}. The logic of such a menu presumes that agents choose correctly. We show that this presumption can fail in a large, high-stakes market: dealers frequently missort, and competitive pressure is part of what corrects it. Whether a screening menu delivers efficient sorting thus depends not only on its design, but on the competitive environment in which agents choose.

Our findings also have implications for competition policy. The case for it usually rests on allocative efficiency---curbing the markups and output restrictions of firms with market power. Our results point to a complementary internal-efficiency channel in the spirit of \citet{leibenstein1966}: competition also improves the quality of firms' decisions. To the extent that antitrust and other pro-competitive policies intensify competitive pressure, they may lead firms to correct costly mistakes of their own---a benefit that conventional analyses of market power overlook. 

A lingering question from our study is what reason or combination of reasons leads dealers to so often choose a contract that is not profit maximizing, and why this tendency is reduced as competition intensifies. While our data are ill-suited to answer this question precisely, we flag several candidates that we believe contribute. Perhaps most likely, in our opinion, is that contract choices are non-profit-maximizing partly due to inattention. Competition may then provide a ``wake-up call'' that reduces inattention and encourages more careful, data-based forecasting from the relevant decision-maker in the firm.  The propensity to choose contracts that sacrifice profits could also reflect other motives. Even in firms using more data-based approaches to choosing the optimal contract, belief biases such as optimism or pessimism could lead to mistakes. Because sales are uncertain at the moment of contract selection, firms' risk-preferences could contribute to firm choice of contract (although firms deviating from risk neutrality is itself typically considered anomalous behavior in competitive markets). Turning toward more squarely ``behavioral'' stories, non-standard risk preferences (and specifically loss aversion) have been demonstrated among auto dealers facing similar decisions \citep{pierce2020negative}, and also could contribute.  Firms that fear they will ``slack'' could value buying into higher bonuses as a partial commitment device. And finally, social pressure and social comparisons also may contribute.\footnote{Anecdotally, we are aware of dealers who perceive there to be pressure from the manufacturer to take the high-bonus tier as it signals the intent to be a bigger player in the market.} In our opinion, all of these potential channels are interesting and disentangling them would be worthwhile; we commend this task to future researchers.

	\clearpage
	\begin{singlespace}
		\bibliographystyle{apa}
		\bibliography{add_dealership.bib}
	\end{singlespace}
	
	\clearpage
	\clearpage
\section*{Figures and Tables}

\clearpage

\clearpage

\begin{table}[H]
	\caption{Summary Statistics}\label{tab:summ}
	{\footnotesize
		\begin{centering} 
			\protect
			\begin{tabular}{lccc}
				\addlinespace \addlinespace \midrule			
				& Mean & SD  & N   \\
				& (1) & (2)  & (3)   \\
				\midrule 	
				\multicolumn{4}{l}{\hspace{-1em} \textbf{Panel A. Dealership Characteristics}} \\ 
				Observed in 2017 &1.000 &0.013 &150,096 \\
Observed in 2018 &0.976 &0.152 &150,096 \\
Observed in 2019 &0.930 &0.255 &150,096 \\
Average number of salespeople &7.9 &6.4 &140,881 \\
Average vehicle sales per month &60.0 &248.8 &143,309 \\
Share of cars in total vehicle sales &0.159 &0.145 &142,543 \\
Share of light duty trucks in total vehicle sales &0.841 &0.145 &142,543 \\
Fraction who left the market in our sample &0.111 &0.314 &150,096 \\
Fraction enrolled in contract at least one year &0.595 &0.491 &150,096 \\
Average number of years enrolled in contract &1.6 &1.4 &150,096 \\
 \midrule
				
				\multicolumn{4}{l}{\hspace{-1em} \textbf{Panel B. Characteristics of Program Participants}} \\ 
				Average number of salespeople &9.6 &7.1 &80,593 \\
Average vehicle sales per month &77.9 &317.6 &81,836 \\
Share of cars in total vehicle sales &0.199 &0.136 &81,702 \\
Share of light duty trucks in total vehicle sales &0.800 &0.136 &81,702 \\
Enrolled in high-bonus contract &0.814 &0.389 &82,318 \\
Fraction months dealers met additional requirements &0.982 &0.133 &81,345 \\
Fraction months dealers received a positive bonus &0.676 &0.468 &81,345 \\
Average yearly bonus received from contract (USD) &298,265.3 &380,247.6 &82,344 \\
Average yearly profit from contract (bonus minus fee) &261,999.6 &377,858.7 &82,332 \\
 \midrule
				
				\multicolumn{4}{l}{\hspace{-1em} \textbf{Panel C. Year-to-Year Enrollment Transitions}} \\ 
				Fraction remained enrolled in contract &0.522 &0.500 &100,068 \\
Fraction remained not enrolled in contract &0.420 &0.494 &100,068 \\
Fraction switched from enrolled to not enrolled &0.018 &0.132 &100,068 \\
Fraction switched from not enrolled to enrolled &0.039 &0.194 &100,068 \\
Fraction remained enrolled in high tier &0.772 &0.419 &54,120 \\
Fraction remained enrolled in low tier &0.114 &0.318 &54,120 \\
Fraction switched from high to low tier &0.011 &0.105 &54,120 \\
Fraction switched from low to high tier &0.069 &0.254 &54,120 \\
 \midrule

				\multicolumn{4}{l}{\hspace{-1em} \textbf{Panel D. Market Competition Measures}} \\ 
				Leave-one-out share of largest competitor &0.214 &0.164 &145,116 \\
Leave-one-out share of top 3 competitors &0.407 &0.210 &145,116 \\
Leave-one-out share of top 5 competitors &0.526 &0.225 &145,116 \\
Herfindahl-Hirschman Index (HHI) &0.120 &0.130 &145,116 \\
 \midrule

			\end{tabular}
			\par\end{centering}
		
		\singlespacing\justify\footnotesize
		\textit{Notes:} This table presents summary statistics for the dealership sample. Panel A reports characteristics for all dealerships. Panel B restricts to dealerships enrolled in the incentive program for at least one year. Panel C shows transition rates between enrollment states across consecutive years. Panel D presents leave-one-out measures of local market competition at the DMA level, each computed excluding the dealership itself: the Herfindahl-Hirschman Index and the combined market share of the largest 1, 3, and 5 competitors.
		
	}
\end{table}

\begin{table}[H]
	\caption{Actual Versus Ex-Post Optimal Contract Choices} \label{tab:ctfl}
	{\footnotesize
		\begin{center}
			\newcommand\w{2.5}
			\begin{tabular}{l@{}lR{\w cm}@{}L{0.45cm}R{\w cm}@{}L{0.45cm}R{\w cm}@{}L{0.45cm}R{\w cm}@{}L{0.45cm}R{\w cm}@{}L{0.45cm}R{\w cm}@{}L{0.45cm}R{\w cm}@{}L{0.45cm}R{\w cm}@{}L{0.45cm}}
				\midrule
				&&  && \multicolumn{4}{c}{By Dealership's Choice:}  \\  \cmidrule{5-8}
				&& All enrolled dealerships && High-bonus contract && Low-bonus contract \\
				&& (1) && (2) && (3)  \\ \midrule					 					  
				\multicolumn{8}{l}{\hspace{-1em}\textbf{Panel A. Ex-Post Suboptimal Choices}} \\ \addlinespace
				Not ex-post max.&&0.201&\sym{***}&0.042&\sym{***}&0.159&\sym{***}\\
&&(0.007&)&(0.003&)&(0.006&)\\

				\addlinespace \midrule									
				\multicolumn{8}{l}{\hspace{-1em}\textbf{Panel B. Distribution of Ex-Post Optimal Choices}} \\ \addlinespace
				
				No enroll&&0.073&\sym{***}&0.034&\sym{***}&0.040&\sym{***}\\
&&(0.004&)&(0.002&)&(0.003&)\\
  
				High tier&&0.119&\sym{***}& &&0.119&\sym{***}\\
&&(0.005&)&&&(0.005&)\\

				Low tier&&0.009&\sym{***}&0.009&\sym{***}& &\\
&&(0.001&)&(0.001&)&&\\

				\midrule
				\(N\) (dealerships)&&2,482&&2,482&&2,482&\\
					
				\(N\) (dealerships$-$years)&&6,855&&6,855&&6,855&\\
	
				\midrule
			\end{tabular}
		\end{center}
		\begin{singlespace}  \vspace{-.5cm}
			
			\noindent \justify \textit{Notes:} This table shows a comparison of actual and ex-post profit-maximizing contract choices. Panel A reports the fraction of dealerships selecting ex-post suboptimal contracts. Panel B shows the distribution of ex-post optimal choices (non-enrollment, high-bonus tier, or low-bonus tier) conditional on actual selection. Heteroskedasticity-robust standard errors clustered at the dealership level in parentheses. \sym{*}~$p<0.10$, \sym{**}~$p<0.05$, \sym{***}~$p<0.01$.
			
		\end{singlespace} 	
	}
\end{table}

\clearpage
\begin{table}[H]{\footnotesize
		\begin{center}
		\caption{Market Concentration and Ex-Post Suboptimal Choices} \label{tab:reg_competition}
			\newcommand\w{1.5}
			\begin{tabular}{l@{}lR{\w cm}@{}L{0.5cm}R{\w cm}@{}L{0.5cm}R{\w cm}@{}L{0.5cm}R{\w cm}@{}L{0.5cm}}
				\midrule
				&& \multicolumn{8}{c}{Outcome: $=1$ if contract choice was ex-post suboptimal} \\ \cmidrule{3-10}
				&& (1) && (2) && (3) && (4)  \\
				\midrule
				\multicolumn{10}{l}{\hspace{-1em}\textbf{Panel A. Year fixed effects}} \\ \addlinespace
				Sales share top 1 dealer&&0.109&\sym{*}&&&&&&\\
&&(0.059&)&&&&&&\\
Sales share top 3 dealers&&&&0.160&\sym{***}&&&&\\
&&&&(0.041&)&&&&\\
Sales share top 5 dealers&&&&&&0.178&\sym{***}&&\\
&&&&&&(0.037&)&&\\
HHI&&&&&&&&0.153&\sym{**}\\
&&&&&&&&(0.072&)\\
Constant&&0.178&\sym{***}&0.137&\sym{***}&0.109&\sym{***}&0.183&\sym{***}\\
&&(0.016&)&(0.019&)&(0.021&)&(0.013&)\\
   \midrule
				\(N\) (dealerships$-$years)&&6,842&&6,842&&6,842&&6,842&\\
 \midrule \addlinespace \addlinespace
				\multicolumn{10}{l}{\hspace{-1em}\textbf{Panel B. Year and dealer fixed effects}} \\ \addlinespace
				Sales share top 1 dealer&&0.176&\sym{**}&&&&&&\\
&&(0.073&)&&&&&&\\
Sales share top 3 dealers&&&&0.153&&&&&\\
&&&&(0.094&)&&&&\\
Sales share top 5 dealers&&&&&&0.253&\sym{**}&&\\
&&&&&&(0.103&)&&\\
HHI&&&&&&&&0.147&\\
&&&&&&&&(0.096&)\\
Constant&&0.155&\sym{***}&0.131&\sym{***}&0.061&&0.175&\sym{***}\\
&&(0.015&)&(0.037&)&(0.053&)&(0.011&)\\
   \midrule
				\(N\) (dealerships$-$years)&&6,675&&6,675&&6,675&&6,675&\\
 \midrule
			\end{tabular}%
		\end{center}
		\begin{singlespace}  \vspace{-.5cm}
			\noindent \justify		
			
			\textit{Notes:} This table displays estimates of the relationship between market concentration and ex-post suboptimal contract choices. 
			
			The dependent variable equals one if a dealership's realized profit under its chosen contract is lower than under the best alternative (see Section~\ref{sub:mistakes} for details). Each column shows the results using a different measure of market concentration. Market share measures are computed as the share of vehicles sold by the largest $n$ dealerships in a DMA-year. The HHI is calculated as the sum of squared market shares of all dealerships in a DMA, with higher values indicating more concentrated markets. All measures are calculated at the Designated Market Area (DMA) level for each year. Panel A includes year fixed effects only; Panel B adds dealer fixed effects.
			
			Heteroskedasticity-robust standard errors clustered at the DMA level in parentheses. \sym{*}~$p<0.10$, \sym{**}~$p<0.05$, \sym{***}~$p<0.01$.

		\end{singlespace}
	}
\end{table}%

\clearpage
\begin{table}[H]{\footnotesize
		\begin{center}
			\caption{Market Concentration and Suboptimal Choices: Non-Exiting Dealerships} \label{tab:reg_competition_noexit}
			\newcommand\w{1.5}
			\begin{tabular}{l@{}lR{\w cm}@{}L{0.5cm}R{\w cm}@{}L{0.5cm}R{\w cm}@{}L{0.5cm}R{\w cm}@{}L{0.5cm}}
				\midrule
				&& \multicolumn{8}{c}{Outcome: $=1$ if contract choice was ex-post suboptimal} \\ \cmidrule{3-10}
				&& (1) && (2) && (3) && (4)  \\
				\midrule
				\multicolumn{10}{l}{\hspace{-1em}\textbf{Panel A. Year fixed effects}} \\ \addlinespace
				Sales share top 1 dealer&&0.114&\sym{*}&&&&&&\\
&&(0.058&)&&&&&&\\
Sales share top 3 dealers&&&&0.162&\sym{***}&&&&\\
&&&&(0.041&)&&&&\\
Sales share top 5 dealers&&&&&&0.179&\sym{***}&&\\
&&&&&&(0.038&)&&\\
HHI&&&&&&&&0.160&\sym{**}\\
&&&&&&&&(0.071&)\\
Constant&&0.165&\sym{***}&0.125&\sym{***}&0.096&\sym{***}&0.170&\sym{***}\\
&&(0.016&)&(0.019&)&(0.021&)&(0.013&)\\
   \midrule
				\(N\) (dealerships$-$years)&&6,449&&6,449&&6,449&&6,449&\\
 \midrule \addlinespace \addlinespace
				\multicolumn{10}{l}{\hspace{-1em}\textbf{Panel B. Year and dealer fixed effects}} \\ \addlinespace
				Sales share top 1 dealer&&0.184&\sym{**}&&&&&&\\
&&(0.073&)&&&&&&\\
Sales share top 3 dealers&&&&0.159&\sym{*}&&&&\\
&&&&(0.090&)&&&&\\
Sales share top 5 dealers&&&&&&0.269&\sym{***}&&\\
&&&&&&(0.101&)&&\\
HHI&&&&&&&&0.160&\sym{*}\\
&&&&&&&&(0.089&)\\
Constant&&0.144&\sym{***}&0.120&\sym{***}&0.044&&0.164&\sym{***}\\
&&(0.015&)&(0.036&)&(0.052&)&(0.010&)\\
   \midrule
				\(N\) (dealerships$-$years)&&6,349&&6,349&&6,349&&6,349&\\
 \midrule
			\end{tabular}%
		\end{center}
		\begin{singlespace}  \vspace{-.5cm}
			\noindent \justify

			\textit{Notes:} This table re-estimates the relationship between market concentration and suboptimal contract choices on the sample of dealerships that never exit during the sample period (excluding the \getval{n_exit_dealers}, or \getval{exit_share_pct} percent, that do). The dependent variable equals one if a dealership's realized annual profit under its chosen contract was lower than under the best alternative. Columns 1--4 use continuous concentration measures: market share of the largest dealer, top 3 dealers, top 5 dealers, and the Herfindahl-Hirschman Index. Panel A includes year fixed effects only; Panel B adds dealer fixed effects. Heteroskedasticity-robust standard errors clustered at the DMA level in parentheses. \sym{*}~$p<0.10$, \sym{**}~$p<0.05$, \sym{***}~$p<0.01$.

		\end{singlespace}
	}
\end{table}%

\clearpage

  \begin{landscape}
  \begin{table}[H]
    \caption{Market concentration and sales predictability}\label{tab:pred_reg}
    \centering
    \footnotesize
    \setlength{\tabcolsep}{3pt}
    \begin{tabular}{l*{9}{c}}
      \toprule
      & \multicolumn{3}{c}{Forecast error} & \multicolumn{3}{c}{Sales variability} & \multicolumn{3}{c}{Relative variability} \\
\cmidrule(lr){2-4}\cmidrule(lr){5-7}\cmidrule(lr){8-10}
& (1) & (2) & (3) & (4) & (5) & (6) & (7) & (8) & (9) \\
\midrule
Top-1 dealer share & $-$0.146 & $-$0.146 & $-$0.242 & $-$0.104 & $-$0.104 & $-$0.054 & 0.036\sym{**} & 0.036\sym{**} & 0.009 \\
 & (0.136) & (0.135) & (0.288) & (0.081) & (0.081) & (0.092) & (0.017) & (0.017) & (0.022) \\
\addlinespace
Top-3 dealer share & $-$0.258\sym{**} & $-$0.258\sym{**} & $-$0.090 & $-$0.146\sym{**} & $-$0.147\sym{**} & 0.029 & 0.039\sym{***} & 0.039\sym{***} & 0.016 \\
 & (0.122) & (0.122) & (0.428) & (0.072) & (0.072) & (0.108) & (0.013) & (0.013) & (0.028) \\
\addlinespace
Top-5 dealer share & $-$0.316\sym{**} & $-$0.316\sym{**} & 0.295 & $-$0.171\sym{**} & $-$0.172\sym{**} & 0.046 & 0.036\sym{***} & 0.036\sym{***} & 0.012 \\
 & (0.127) & (0.127) & (0.492) & (0.074) & (0.074) & (0.131) & (0.012) & (0.012) & (0.033) \\
\addlinespace
HHI & $-$0.251\sym{*} & $-$0.252\sym{*} & $-$0.327 & $-$0.152\sym{*} & $-$0.153\sym{*} & $-$0.061 & 0.041\sym{**} & 0.041\sym{**} & $-$0.010 \\
 & (0.139) & (0.139) & (0.245) & (0.085) & (0.085) & (0.098) & (0.019) & (0.019) & (0.028) \\
\midrule
Year FE   & No & Yes & Yes & No & Yes & Yes & No & Yes & Yes \\
Dealer FE & No & No  & Yes & No & No  & Yes & No & No  & Yes \\
\(N\) & 7,928 & 7,928 & 7,730 & 12,093 & 12,093 & 11,997 & 12,093 & 12,093 & 11,997 \\

      \bottomrule
    \end{tabular}

    {\footnotesize \singlespacing \justify \textit{Notes:} Each cell is a separate regression of the column outcome on one concentration measure (top-1/3/5 dealer share or HHI); Heteroskedasticity-robust standard errors clustered by DMA in parentheses; all outcomes winsorized at the 1st and 99th percentiles. The three dealer-year outcomes are: \textit{Forecast error}, the squared residual from regressing a dealer's annual sales on its prior-year sales (in thousands of units$^2$); \textit{Sales variability}, the squared deviation of annual sales from the dealer's own mean, $(x-\bar x)^2$ (in thousands); and \textit{Relative variability}, the mean-scaled version $((x-\bar x)/\bar x)^2$ (whose dealer-level average is the squared coefficient of variation). Specifications (2) and (3) add year and then year-and-dealer fixed effects. \sym{*}~$p<0.10$, \sym{**}~$p<0.05$, \sym{***}~$p<0.01$. \par}
  \end{table}
  \end{landscape}

	\clearpage
\clearpage 
\appendix
\begin{center}\noindent {\LARGE \textbf{Appendix}}\end{center}
\label{app:figs}

\setcounter{table}{0}
\setcounter{figure}{0}
\setcounter{equation}{0}	
\renewcommand{\thetable}{A\arabic{table}}
\renewcommand{\thefigure}{A\arabic{figure}}
\renewcommand{\theequation}{A\arabic{equation}}

\section{Appendix Figures and Tables}

\begin{figure}[H]
	\caption{Ex-Post Suboptimal Contract Choices by Dealership Size}\label{fig:mistake_size}
	\centering
	\includegraphics[width=.75\linewidth]{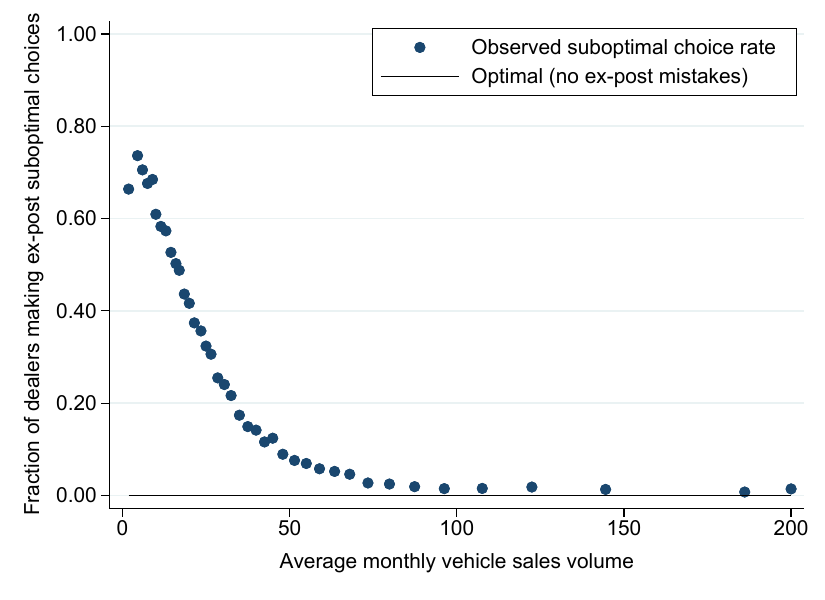}
		
	{\footnotesize
		\singlespacing \justify
		
		\textit{Notes:} This figure shows the relationship between dealership size and ex-post suboptimal contract choices. Binned scatterplot (40 equal-sized bins) shows the fraction of dealerships making suboptimal choices against average monthly vehicle sales. A choice is suboptimal if realized profit under the chosen contract is lower than under the best alternative. The horizontal line at zero represents perfect optimization where all dealerships make profit-maximizing decisions.
		
	}
	
\end{figure}

\clearpage

  \begin{figure}[H]
    \caption{Year-over-year change in annual dealership sales}\label{fig:yoy_sales}
    \centering
    \begin{subfigure}[t]{.48\textwidth}
      \caption*{Panel A. Change in units}
      \centering
      \includegraphics[width=\linewidth]{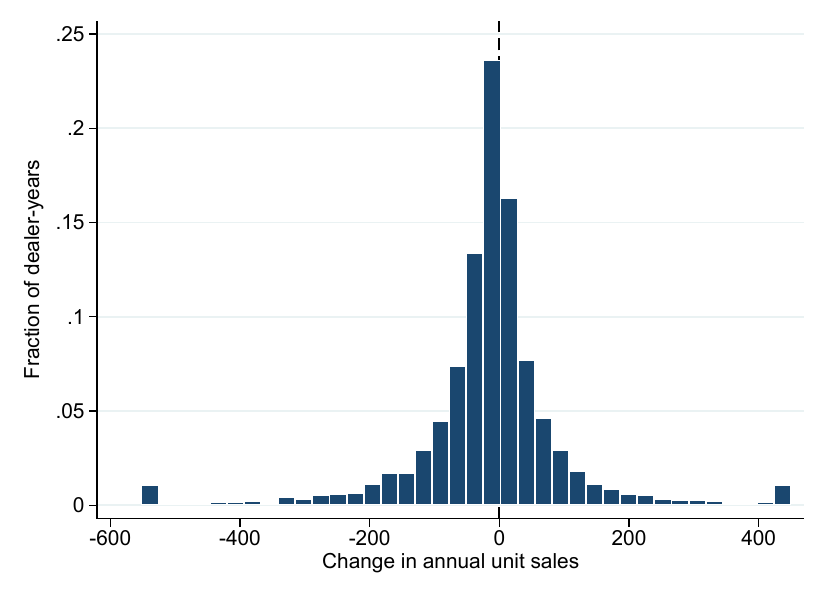}
    \end{subfigure}
    \hfill
    \begin{subfigure}[t]{0.48\textwidth}
      \caption*{Panel B. Percent change}
      \centering
      \includegraphics[width=\linewidth]{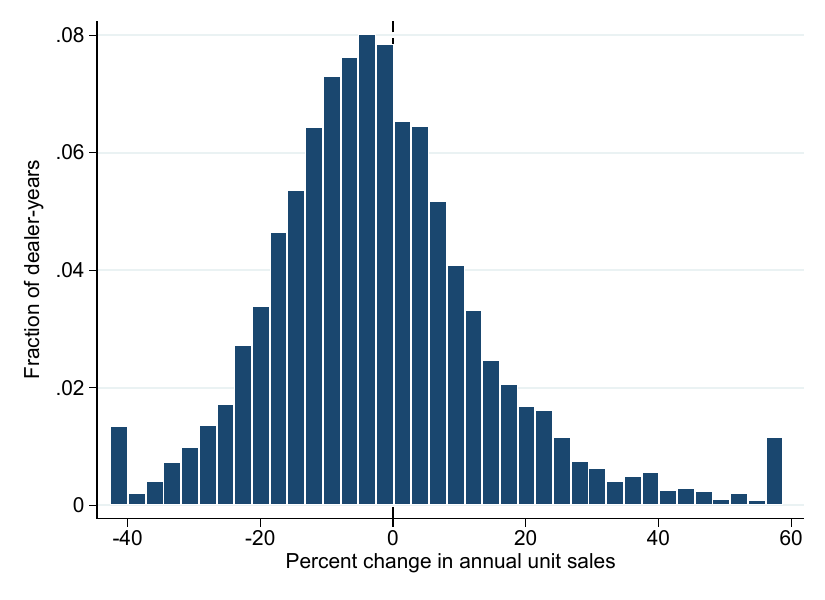}
    \end{subfigure}
    \\[0.5em]
    {\footnotesize
      \singlespacing \justify
      \textit{Notes:} Distribution of the year-over-year change in a dealer's annual unit sales, pooling the 2017$\to$2018 and 2018$\to$2019 transitions (\getval{yoy_n_dealeryears} dealer-years, \getval{yoy_n_dealers} dealers, incumbent balanced panel). Panel A is the change in units; Panel B the percent change. Both winsorized at the 1st and 99th percentiles (units: $[\getval{yoy_units_lo},+\getval{yoy_units_hi}]$; percent: $[\getval{yoy_pct_lo},+\getval{yoy_pct_hi}]$). Dashed line marks zero. \par
    }
  \end{figure}

\clearpage

\foreach \year in {2017, 2018, 2019} {
\clearpage
\begin{figure}[H]
\caption{Market Concentration and Ex-Post Suboptimal Choices (\year)}\label{fig:binsc_comp_\year}
	\centering
	\begin{subfigure}[t]{.48\textwidth}
		\caption*{Panel A. Market Share of \\ Largest Dealer}
		\centering
		\includegraphics[width=\textwidth]{results/binsc_shr_sales_top1_\year}
	\end{subfigure}
	\hfill
	\begin{subfigure}[t]{.48\textwidth}
		\caption*{Panel B. Market Share of \\ Top 3 Dealers}
		\centering
		\includegraphics[width=\textwidth]{results/binsc_shr_sales_top3_\year}
	\end{subfigure}
	\hfill
	\begin{subfigure}[t]{.48\textwidth}
		\caption*{Panel C. Market Share of \\ Top 5 Dealers}
		\centering
		\includegraphics[width=\textwidth]{results/binsc_shr_sales_top5_\year}
	\end{subfigure}
	\hfill
	\begin{subfigure}[t]{.48\textwidth}
		\caption*{Panel D. Herfindahl-Hirschman \\ Index (HHI)}
		\centering
		\includegraphics[width=\textwidth]{results/binsc_hhi_\year}
	\end{subfigure}

	{\footnotesize
		\singlespacing \justify

		\textit{Notes:} This figure replicates Figure~\ref{fig:binsc_comp} using only \year\ data. Binned scatterplots (20 equal-sized bins) with 95 percent confidence intervals show the fraction of dealerships making ex-post suboptimal tier choices against each concentration measure, with each dealership entering as a single observation in \year. An ex-post mistake occurs when a dealership's realized profit under its chosen contract is lower than under the best alternative. The maroon line shows the best linear fit.

	}

\end{figure}
}

\foreach \year in {2017, 2018, 2019} {
\clearpage
	\begin{table}[H]
	\caption{Actual versus Ex-Post Optimal Contract Choices in \year} \label{tab:ctfl_\year}	
		{\footnotesize
			
			\begin{center}
				\newcommand\w{2}
				\begin{tabular}{l@{}lR{\w cm}@{}L{0.45cm}R{\w cm}@{}L{0.45cm}R{\w cm}@{}L{0.45cm}R{\w cm}@{}L{0.45cm}R{\w cm}@{}L{0.45cm}R{\w cm}@{}L{0.45cm}R{\w cm}@{}L{0.45cm}}
					\midrule
					&&  && \multicolumn{4}{c}{By Dealership's Choice:}  \\  \cmidrule{5-8}
					&& All enrolled dealerships && High-bonus contract && Low-bonus contract \\
					&& (1) && (2) && (3)  \\ \midrule                                          
					\multicolumn{8}{l}{\hspace{-1em}\textbf{Panel A. Ex-Post Suboptimal Choices}} \\ \addlinespace
					\input{results/shr_mistake\year.tex}  
					
					\addlinespace \midrule                                    
					\multicolumn{8}{l}{\hspace{-1em}\textbf{Panel B. Distribution of Ex-Post Optimal Choices}} \\ \addlinespace
					
					\input{results/shr_noenroll\year.tex}  
					\input{results/shr_high\year.tex}
					\input{results/shr_low\year.tex}
					
					\midrule
					\input{results/shr_mistake_N\year.tex}                    
					
					\midrule
				\end{tabular}
			\end{center}
			\begin{singlespace}  \vspace{-.5cm}
				
				\noindent \justify \textit{Notes:} This table is a replication of Table~\ref{tab:ctfl} using only \year\ data. The table compares dealerships' actual tier choices with ex-post profit-maximizing alternatives. Heteroskedasticity-robust standard errors clustered at the dealership level in parentheses. \sym{*}~$p<0.10$, \sym{**}~$p<0.05$, \sym{***}~$p<0.01$.

			\end{singlespace}    
		}
	\end{table}
	
}

\begin{table}[H]{\footnotesize
		\begin{center}
			\caption{Market Concentration and Contract Choice: Robustness to Outliers (Excluding DMAs More Than 2 SD from the Mean)} \label{tab:reg_competition_out2sd}
			\newcommand\w{1.5}
			\begin{tabular}{l@{}lR{\w cm}@{}L{0.5cm}R{\w cm}@{}L{0.5cm}R{\w cm}@{}L{0.5cm}R{\w cm}@{}L{0.5cm}}
				\midrule
				&& \multicolumn{8}{c}{Outcome: $=1$ if contract choice was ex-post suboptimal} \\ \cmidrule{3-10}
				&& (1) && (2) && (3) && (4)   \\
				\midrule
				\multicolumn{10}{l}{\hspace{-1em}\textbf{Panel A. Year fixed effects}} \\ \addlinespace
				Sales share top 1 dealer&&0.202&\sym{**}&&&&&&\\
&&(0.080&)&&&&&&\\
Sales share top 3 dealers&&&&0.185&\sym{***}&&&&\\
&&&&(0.055&)&&&&\\
Sales share top 5 dealers&&&&&&0.183&\sym{***}&&\\
&&&&&&(0.045&)&&\\
HHI&&&&&&&&0.489&\sym{***}\\
&&&&&&&&(0.135&)\\
Constant&&0.163&\sym{***}&0.129&\sym{***}&0.107&\sym{***}&0.155&\sym{***}\\
&&(0.017&)&(0.021&)&(0.023&)&(0.015&)\\
   \midrule
				\(N\) (dealerships$-$years)&&6,491&&6,535&&6,558&&6,482&\\
 \midrule \addlinespace\addlinespace
				
				\multicolumn{10}{l}{\hspace{-1em}\textbf{Panel B. Year and dealer fixed effects}} \\ \addlinespace
				Sales share top 1 dealer&&0.179&\sym{*}&&&&&&\\
&&(0.108&)&&&&&&\\
Sales share top 3 dealers&&&&0.172&\sym{*}&&&&\\
&&&&(0.094&)&&&&\\
Sales share top 5 dealers&&&&&&0.250&\sym{**}&&\\
&&&&&&(0.103&)&&\\
HHI&&&&&&&&0.181&\\
&&&&&&&&(0.205&)\\
Constant&&0.159&\sym{***}&0.124&\sym{***}&0.065&&0.175&\sym{***}\\
&&(0.019&)&(0.035&)&(0.051&)&(0.019&)\\
   \midrule
				\(N\) (dealerships$-$years)&&6,297&&6,375&&6,397&&6,296&\\
 \midrule
				
			\end{tabular}%
		\end{center}
		\begin{singlespace}  \vspace{-.5cm}
			\noindent \justify
			
			\textit{Notes:} This table reproduces Table~\ref{tab:reg_competition} excluding outlier DMAs whose concentration measure lies more than 2 standard deviations from its mean (applied separately for each of the four measures). The dependent variable equals one if a dealership's realized profit under its chosen contract is lower than under the best alternative. Each column uses a different concentration measure. Panel A includes year fixed effects only; Panel B adds dealer fixed effects. Heteroskedasticity-robust standard errors clustered at the DMA level in parentheses. \sym{*}~$p<0.10$, \sym{**}~$p<0.05$, \sym{***}~$p<0.01$.
			
		\end{singlespace} 	
	}
\end{table}%

\clearpage

\begin{table}[H]{\footnotesize
		\begin{center}
			\caption{Market Concentration and Contract Choice: Robustness to Outliers (Excluding DMAs More Than 3 SD from the Mean)} \label{tab:reg_competition_out3sd}
			\newcommand\w{1.5}
			\begin{tabular}{l@{}lR{\w cm}@{}L{0.5cm}R{\w cm}@{}L{0.5cm}R{\w cm}@{}L{0.5cm}R{\w cm}@{}L{0.5cm}}
				\midrule
				&& \multicolumn{8}{c}{Outcome: $=1$ if contract choice was ex-post suboptimal} \\ \cmidrule{3-10}
				&& (1) && (2) && (3) && (4)   \\
				\midrule
				\multicolumn{10}{l}{\hspace{-1em}\textbf{Panel A. Year fixed effects}} \\ \addlinespace
				Sales share top 1 dealer&&0.160&\sym{**}&&&&&&\\
&&(0.073&)&&&&&&\\
Sales share top 3 dealers&&&&0.160&\sym{***}&&&&\\
&&&&(0.041&)&&&&\\
Sales share top 5 dealers&&&&&&0.178&\sym{***}&&\\
&&&&&&(0.037&)&&\\
HHI&&&&&&&&0.290&\sym{**}\\
&&&&&&&&(0.117&)\\
Constant&&0.170&\sym{***}&0.137&\sym{***}&0.109&\sym{***}&0.171&\sym{***}\\
&&(0.017&)&(0.019&)&(0.021&)&(0.015&)\\
   \midrule
				\(N\) (dealerships$-$years)&&6,646&&6,842&&6,842&&6,628&\\
 \midrule \addlinespace\addlinespace
				\multicolumn{10}{l}{\hspace{-1em}\textbf{Panel B. Year and dealer fixed effects}} \\ \addlinespace
				Sales share top 1 dealer&&0.190&\sym{**}&&&&&&\\
&&(0.080&)&&&&&&\\
Sales share top 3 dealers&&&&0.153&&&&&\\
&&&&(0.094&)&&&&\\
Sales share top 5 dealers&&&&&&0.253&\sym{**}&&\\
&&&&&&(0.103&)&&\\
HHI&&&&&&&&0.187&\\
&&&&&&&&(0.125&)\\
Constant&&0.156&\sym{***}&0.131&\sym{***}&0.061&&0.173&\sym{***}\\
&&(0.015&)&(0.037&)&(0.053&)&(0.012&)\\
   \midrule
				\(N\) (dealerships$-$years)&&6,449&&6,675&&6,675&&6,436&\\
 \midrule
			\end{tabular}%
		\end{center}
		\begin{singlespace}  \vspace{-.5cm}
			\noindent \justify
			\textit{Notes:} This table reproduces Table~\ref{tab:reg_competition} excluding outlier DMAs whose concentration measure lies more than 3 standard deviations from its mean (applied separately for each of the four measures). The dependent variable equals one if a dealership's realized profit under its chosen contract is lower than under the best alternative. Each column uses a different concentration measure. Panel A includes year fixed effects only; Panel B adds dealer fixed effects. Heteroskedasticity-robust standard errors clustered at the DMA level in parentheses. \sym{*}~$p<0.10$, \sym{**}~$p<0.05$, \sym{***}~$p<0.01$.
		\end{singlespace}
	}
\end{table}%

\begin{table}[H]{\footnotesize
    \begin{center}
      \caption{Market Concentration and Dealership Exit} \label{tab:exit}
      \newcommand\w{1.5}
      \begin{tabular}{l@{}lR{\w cm}@{}L{0.5cm}R{\w cm}@{}L{0.5cm}R{\w cm}@{}L{0.5cm}R{\w cm}@{}L{0.5cm}}
        \midrule
        && \multicolumn{8}{c}{Outcome: $=1$ if the dealership exits the sample} \\ \cmidrule{3-10}
        && (1) && (2) && (3) && (4)  \\
        \midrule
        \multicolumn{10}{l}{\hspace{-1em}\textbf{Panel A. Year fixed effects}} \\ \addlinespace
        Sales share top 1 dealer&&$-$0.002&&&&&&&\\
&&(0.010&)&&&&&&\\
Sales share top 3 dealers&&&&0.002&&&&&\\
&&&&(0.009&)&&&&\\
Sales share top 5 dealers&&&&&&0.002&&&\\
&&&&&&(0.008&)&&\\
HHI&&&&&&&&$-$0.006&\\
&&&&&&&&(0.012&)\\
Constant&&0.039&\sym{***}&0.037&\sym{***}&0.037&\sym{***}&0.039&\sym{***}\\
&&(0.003&)&(0.004&)&(0.005&)&(0.002&)\\
   \midrule
        \(N\) (dealerships$-$years)&&12,093&&12,093&&12,093&&12,093&\\
 \midrule \addlinespace \addlinespace
        \multicolumn{10}{l}{\hspace{-1em}\textbf{Panel B. Year and dealer fixed effects}} \\ \addlinespace
        Sales share top 1 dealer&&$-$0.011&&&&&&&\\
&&(0.031&)&&&&&&\\
Sales share top 3 dealers&&&&0.035&&&&&\\
&&&&(0.042&)&&&&\\
Sales share top 5 dealers&&&&&&0.015&&&\\
&&&&&&(0.048&)&&\\
HHI&&&&&&&&0.012&\\
&&&&&&&&(0.048&)\\
Constant&&0.033&\sym{***}&0.016&&0.022&&0.029&\sym{***}\\
&&(0.007&)&(0.017&)&(0.025&)&(0.006&)\\
   \midrule
        \(N\) (dealerships$-$years)&&11,997&&11,997&&11,997&&11,997&\\
 \midrule
      \end{tabular}%
    \end{center}
    \begin{singlespace} \vspace{-.5cm}
      \noindent \justify
      \textit{Notes:} The dependent variable equals one in a dealership's last observed year if it exits the sample before the end of our window, and zero in the dealer-years it is present. Each column uses a different measure of local market concentration, computed at the DMA-year level as in the paper's main concentration table. Panel A includes year fixed effects; Panel B adds dealer fixed effects. Because exit is a rare, one-time event---each exiting dealer contributes a single zero-to-one transition in its final year, and non-exiting dealers are all zeros---there is little within-dealer variation to exploit, so the dealer fixed effects in Panel B are underpowered. Heteroskedasticity-robust standard errors clustered at the DMA level in parentheses. \sym{*}~$p<0.10$, \sym{**}~$p<0.05$, \sym{***}~$p<0.01$.
    \end{singlespace}
  }
\end{table}

\end{document}